\documentclass[3p,review]{elsarticle}
\usepackage{bm}
\usepackage{graphicx}
\usepackage{color}
\usepackage{subfigure}
\usepackage{latexsym}
\usepackage{amsthm}
\usepackage{amssymb}
\usepackage{amsmath}
\usepackage{wrapfig}
\usepackage{ulem}
\biboptions{sort&compress}
\usepackage{amsmath,amsfonts,amssymb,bm,cancel}
\usepackage{soul,cancel}
\usepackage{epsf,amsmath,amssymb,verbatim,color,multirow,pifont,graphicx,cleveref,tabularx}

\usepackage{bm}
\usepackage{xcolor}

\def\bea{\begin{eqnarray}}
\def\eea{\end{eqnarray}}
\newcommand{\Dirac}{{\bm{\mathcal{D}}} \hskip -2.2mm \slash}

\newtheorem{theorem}{Theorem}
\newtheorem{remark}[theorem]{Remark}

\begin{document}

\title{The three way Dirac operator and \\ dynamical Turing and Dirac induced patterns on nodes and links}                      



%
\author{Riccardo Muolo}



\address{{Department of Systems and Control Engineering, Tokyo Institute of Technology},\\
    {2 Chome-12-1 Ookayama}, 
    {Meguro-ku, Tokyo},
    {152-8552}, 
   {Japan}}
    
    \address{{Department of Mathematics \& naXys, Namur Institute for Complex Systems, Université de Namur},\\
   {Rue Grafé 2}, 
   {Namur},
   {5000}, 
    {Belgium}}

\author{Timoteo Carletti}
\address{{Department of Mathematics \& naXys, Namur Institute for Complex Systems, Université de Namur},\\
   {Rue Grafé 2}, 
   {Namur},
   {5000}, 
    {Belgium}}

\author{Ginestra Bianconi}

\ead{ginestra.bianconi@gmail.com}

    \address{{Centre for Complexity Science, School of Mathematical Sciences, Queen Mary University of London},\\
    {Mile End Road}, 
    {London},
   {E1 4NS}, 
    {UK}}
    
 \address{{The  Alan  Turing  Institute},
    {The British Library, 96  Euston  Road}, 
    {London},
    {NW1  2DB}, 
    {UK}}



\begin{abstract}
Topological signals are dynamical variables not only defined on nodes but also on links of a network that are gaining significant attention in non-linear dynamics and topology and have important applications in brain dynamics. Here we show that topological signals on nodes and links of a network can generate dynamical patterns when coupled together. In particular, dynamical patterns require at least three topological signals, here taken to be two node signals and one link signal.
In order to couple these signals, we formulate the 3-way topological Dirac operator that generalizes previous definitions of the 2-way and 4-way topological Dirac operators. We characterize the spectral properties of the 3-way Dirac operator and we investigate the dynamical properties of the resulting Turing and Dirac induced patterns. Here we emphasize the distinct dynamical properties of  the Dirac induced patterns which involve topological signals only coupled by the 3-way topological Dirac operator in absence of the Hodge-Laplacian coupling. While the observed Turing patterns generalize the Turing patterns typically investigated on networks, the Dirac induced patterns have no equivalence within the framework of node based Turing patterns.  These results open new scenarios in the study of Turing patterns with possible application to neuroscience and more generally to the study of emergent patterns in complex systems.  
\end{abstract}


\begin{keyword}
Higher-order networks \sep Turing patterns  \sep Topological signals \sep Dirac operator


\end{keyword}
\maketitle

\section{Introduction}
Networks \cite{barabasi,newman2018networks} represent the discrete architecture of complex systems by encoding the set of  interactions (links) existing among their constituent elements (nodes). Network science  is based on the fundamental assumption that the network topology encodes important information about the network dynamics and hence about the function of the underlying complex interacting system.
The statistical and combinatorial properties of the network structure have been shown to be key in shaping the phase diagram of dynamical processes ranging from epidemics to percolation \cite{barrat2008dynamical,dorogovtsev2008critical}. Yet, we are still far from a complete understanding of the interplay between network topology and dynamics that would be necessary, for instance, to transform our understanding of brain dynamics. 

Until recently, the dynamical state of a network has been exclusively described following a node centered approach, where the dynamical variables are exclusively defined on the nodes of the networks.
Nowadays,  it is increasingly recognized that this approach has important limitations and that in a number of cases the description of  network dynamics should include higher-order topological signals, i.e., dynamical variables not only defined on nodes but also on links. Examples of link signals include synaptic signals between neurons or brain link signals at the level of brain regions \cite{santoro2022unveiling,sporns_edge}  and in general currents in biological transportation networks. Moreover, on higher-order networks one can define topological signals also on triangles, tetrahedra, and so on. This shift of perspective significantly enhances  the topological characterization of their structure\cite{giusti2015clique,reimann2017cliques,petri2013topological,lee2021homological,bobrowski2022homological,santos2019topological} and the node-centric investigation of their  dynamics  \cite{mulas_msf,skardal2020higher,gambuzza2018distributed,gambuzza2021stability,de2021phase}.

The study of topological signals requires to combine topology with non-linear dynamics and it is emerging as a field that can  transform our understanding of the interplay between structure and dynamics on simple and higher-order networks \cite{bianconi2021higher,natphys}.
Interestingly, it has been found that topological signals can undergo collective critical phenomena such as topological higher-order Kuramoto model \cite{millan2020explosive,ghorbanchian2021higher,deville2021consensus,petri_hodgeg},  global topological synchronization \cite{carletti2023global} and  higher-order diffusion \cite{torres2020simplicial,ziegler2022balanced,schaub2020random}.  Moreover, real-world topological signals can be predicted and processed with topological  machine learning algorithms \cite{battiloro2023generalized,barbarossa2020topological,schaub2021signal,ebli2020simplicial,hensel2021survey}.

On a simplicial complex, topological signals are modeled, treated and processed by combining non-linear dynamics with algebraic topology and discrete calculus inherited by the simplicial support (which reduces to a network if the simplicial complex is $1$-dimensional). In particular, topological signals of a given dimension are often treated by using the Hodge Laplacian \cite{bianconi2021higher,horak2013spectra} that describes diffusion from $n$-dimensional simplices to $n$-dimensional simplices. 

Recently, the discrete topological Dirac operator \cite{bianconi2021topological} has been shown to be the most suitable operator in situations in which topological signals of different dimensions interact and cross-talk; indeed the Dirac operator allows to project the signal on $n$-dimensional signal to one dimension up or down. For instance in a network the Dirac operator can be used to project node signals to link signals and vice versa.

The Dirac operator has been originally defined in non-commutative geometry and in quantum graphs \cite{post2009first},  but only recently it has been show to lead to topological field theories \cite{bianconi2021topological,bianconi2022dirac,bianconi2023mass} and to be  a key operator in complex systems. Indeed, it is key to treat dynamics of coupled topological signals  \cite{calmon2022dirac,calmon2023local,turing_topological,calmon2023dirac,battiloro2023generalized,nurisso2023unified} on top of having important application to topological analysis of real data \cite{lloyd2016quantum,wee2023persistent}.

In this work, we combine topology and non-linear dynamics to study dynamical pattern formation of topological signals of nodes and links, induced by the Dirac operator.

On continuous domains, reaction-diffusion equations are partial differential equations that can be used to describe chemical reaction systems, ecosystems, neuronal dynamics and fluid-dynamics, and are a core subject of non-linear dynamics \cite{Murray2001}.
Turing patterns are spatial patterns emerging from a diffusion-driven instability of the homogeneous equilibrium state of a reaction-diffusion system and are pivotal to describe pattern formation in the natural world. In a nutshell, a system of two interacting species is perturbed about its spatially homogeneous stable equilibrium, which becomes unstable due to diffusion, giving rise to the celebrated Turing instability from which patterns may originate~\cite{Turing}. The original framework in which Turing conceived this pattern formation mechanism was morphogenesis, but, nowadays, it finds applications in many different field, including biology \cite{kondo2009}, neuroscience \cite{byrne2019next} and even quantum mechanics \cite{kato_nakao_quantum_Turing} and nano-materials \cite{turing_nano}.
Turing theory has been extended on regular lattices in the 70s by Othmer and Scriven \cite{OS1971,OthmerScriven74}, but it is only recently that increasing attention has been devoted to characterize Turing patterns on networks, starting from the seminal work of Nakao and Mikhailov \cite{NM2010}. Turing patterns have been thoroughly studied on different networks topologies, such as directed \cite{Asllani1}, non-normal \cite{jtb} and geometric \cite{van_der_kolk_PRX_2023} ones, with applications ranging from ecology \cite{hata_dispersal} to control \cite{control_hata,control_holme,control_frasca}, to name a few. Moreover, they have also been observed on multilayer \cite{Asllani2014,Asllani2016,KHDG}, temporal \cite{PABFC2017,vangorder2,carletti_fanelli_temporalCSF} and higher-order \cite{carletti2020dynamical,muologallo,gao_turing_simplicial} networks.

In all  these approaches it is  assumed that the dynamics is only localized on the nodes of the network. In order to address this limitation, recently Giambagli et al. \cite{turing_topological} have characterized the formation of topological Turing patterns on nodes and links of networks. In this setting, the dynamical state of the network is encoded by one topological signal (species) on the nodes and one topological signal (species) on the links of the network coupled together by the Dirac operator. 
The topological Turing patterns that emerge in this scenario are inhomogeneous on nodes and links, however they are stationary, i.e., they characterize a heterogeneous static asymptotic state of the network.

In this work, we build a mathematical framework that is able to generate dynamical Turing patterns on nodes and links. The key element of the model is the assumption that the system state is encoded in three different types of topological signals (species): two anchored to nodes and one on links.
In order to treat these topological signals, here we define  the 3-way Dirac operator and its associated gamma matrix. The Dirac operator projects the two nodes signals into the links and the single link signal into the nodes, while the gamma matrix is used to compress the projected nodes signals into a single link signal and to expand the single projected link signal into two nodes signals.
The 3-way Dirac operator couples nodes and links topological signals and is shown here to induce dynamical Turing patterns of topological nodes and links signals displaying very rich and non-trivial dynamical properties.

\section{Introduction to discrete exterior calculus and to the topological Dirac operator}
We consider a network $G=(V,E)$  comprising a set $V$ of $N_{0}=|V|$ nodes, and a set $E$ of $N_{1}=|E|$ links. According to the usual algebraic topology setting links are undirected, yet they are characterized by an orientation from one endnode to the other.
Assuming that the network dynamics is captured by a topological signal on the nodes and a topological signal on the links, the full dynamical state of the network is encoded by the topological spinor $\bm\Phi\in C^{0}\oplus C^{1}$ given by 
\bea
{\mathbf{\Phi}}=\binom{u}{w}\,,
\eea
where $u\in C^1$ is $0$-cochain here encoded by a vector taking real values on each node of the network and $v\in C^1$ is a $1$-cochain here encoded by a vector taking real values on each link of the network, i.e.
\bea
u=\left(\begin{array}{c}
u_1 \\
u_2\\
\vdots\\
u_{N_0}
\end{array}\right) \text{ and  } w=\left(\begin{array}{c}
w_1 \\
w_2\\
\vdots\\
w_{N_1}
\end{array}\right). 
\eea
The discrete exterior calculus is a branch of mathematics that allows to define the discrete gradient and the discrete divergence of these topological signals.
In particular the discrete gradient $\delta_1:C^{0}\to C^{1}$ is the linear operator that acts on topological nodes signals (e.g., the signal $u$) and provides a link topological signal (e.g., $g$ with $g=\delta_1u$); in the considered example we have  
\bea
g_{rs}=u_{s}-u_r \, , \text{for all link $[r,s]$.}
\eea
Moreover the discrete divergence $\delta_1^{\star}:C^{1}\to C^{2}$ is the linear operator that acts on link topological signals  (e.g., $w$) and provides a node topological signal (e.g., $f$ with $f=\delta_1^{\star}w$) where in the considered case we have for all node $r$
\bea
f_{r}=\sum_{\ell\in E^{+}_r}w_{\ell}-\sum_{\ell\in E^{-}_r}w_{\ell}\, ,
\eea
where $E^{+}_r$ is the set of links oriented toward node $r$ while $E^{-}_r$ is the set of links oriented from node $r$ toward their other endnode.
Therefore the discrete divergence associates to a node $r$ the difference between the outward and inward flow from/to the node.
The two linear operators can be represented by the $N_0\times N_1$ boundary matrix ${\bf B}$ whose elements are given by 
 \bea
{\bf B}_{r \ell }=\begin{cases}
-1 ~\mbox{ if } ~\ell=[r,s],\nonumber \\
~1  ~\mbox{ if }~ \ell=[s,r].
\end{cases}  
\eea
In particular we have ${f}={\bf B}w$ and $g={\bf B}^{\top}u$.
On a network the Hodge Laplacians ${\bf L}_{[0]}$ and ${\bf L}_{[1]}$ describe diffusion from nodes to nodes passing through links and from links to links passing through nodes. Therefore the Hodge Laplacians of a network can be build by contracting the boundary matrix in the two possible ways, leading to  ${\bf L}_{[0]}={\bf B}{\bf B}^{\top}$ and ${\bf L}_{[1]}={\bf B}^{\top}{\bf B}$.
The most simple definition of the Hodge-Dirac operator \cite{post2009first,bianconi2021topological} on a network is the operator $\partial:C^{0}\oplus C^1\to C^0\oplus C^1$ that maps topological spinors to topological spinors and is defined as $\partial=\delta_1^{\star}\oplus \delta_1$.
In particular if $\bm\Phi=u\oplus w$ we have 
\bea
\partial (u\oplus w)=\delta_1^{\star} w\oplus \delta_1 u\, ,
\eea
or in matrix form we have
\bea
\bm{\partial}=\left(\begin{array}{cc}
0  & \mathbf{B} \\
\mathbf{B}^{\top}&0 
\end{array}\right)\, .
\eea
The Hodge-Dirac operator can be considered as the ``square root" of the higher-order Laplacian $\boldsymbol{\mathcal{L}}$. Indeed we have 
\bea
\label{eq:delta2L}
\bm{\partial}^2=\boldsymbol{\mathcal{L}}=\left(\begin{array}{cc}
 \mathbf{L}_{[0]}&0 \\
0&\mathbf{L}_{[1]}
\end{array}\right)\, .
\eea
Therefore the non-zero eigenvalues of the Dirac operator are given by plus or minus the square root of  the non-zero eigenvalues of ${\bf L}_{[0]}$ (which is by the way isospectral to ${\bf L}_{[1]}$ on a generic network).
The Hodge-Dirac operator admits eigenvectors that are either harmonic (i.e., associated to the zero eigenvalue of the Hodge-Dirac operator) or chiral.
Here by chiral eigenvectors we refer to the relations of eigenvectors associated with non-zero eigenvalues with the same absolute value. Indeed it can be easily proved that  $\bm\partial$ anticommutes (i.e., $\{\bm\partial,\bm\gamma_0\}=0$) with $\bm\gamma_0$,  where 
\bea
\bm\gamma_0=\left(\begin{array}{cc}
{\bf I}_{N_0}&{ 0}\\
{ 0}&-{\bf I}_{N_1}
\end{array}\right)\, ,
\eea 
where ${\bf I}_X$ indicates the $X\times X$ identity matrix.
This result implies that if $(u,w)^{\top}$ is an eigenvector of $\bm\partial$ with eigenvalue $\hat{\lambda}$, then $(u,-w)^{\top}$ is an eigenvector of $\bm\partial$ with eigenvalue $-\hat\lambda$. This relation between the non-harmonic eigenvectors of the Hodge-Dirac operator is called chirality.
This Hodge-Dirac operator has been used in Ref.\cite{turing_topological} to model Turing patterns on nodes and links. However, by using a topological spinor formed by a single topological nodes signal and a single topological links signal, can only account for stationary Turing patterns.

Interestingly the Hodge-Dirac operator can be coupled with group operations enforced by the so called gamma matrices \cite{bianconi2021topological}. For instance on a lattice the gamma matrices allows to distinguish between gradient and divergence performed along different directions. In particular on a three dimensional lattice in order to distinguish between the $x$, $y$, and $z$ directions one needs to consider two topological signals on the nodes and two topological signals on the links and express the gamma matrices in terms of the Pauli matrices acting on the two dimensional node (or link) signal.
However no existing approach is able to jointly treat and process an odd number of topological signals such as two signal on the nodes and one signal on the links or viceversa two signals on the links and on signal on the nodes. 
\begin{figure}[ht!]
\centering
\includegraphics[width=1\columnwidth]{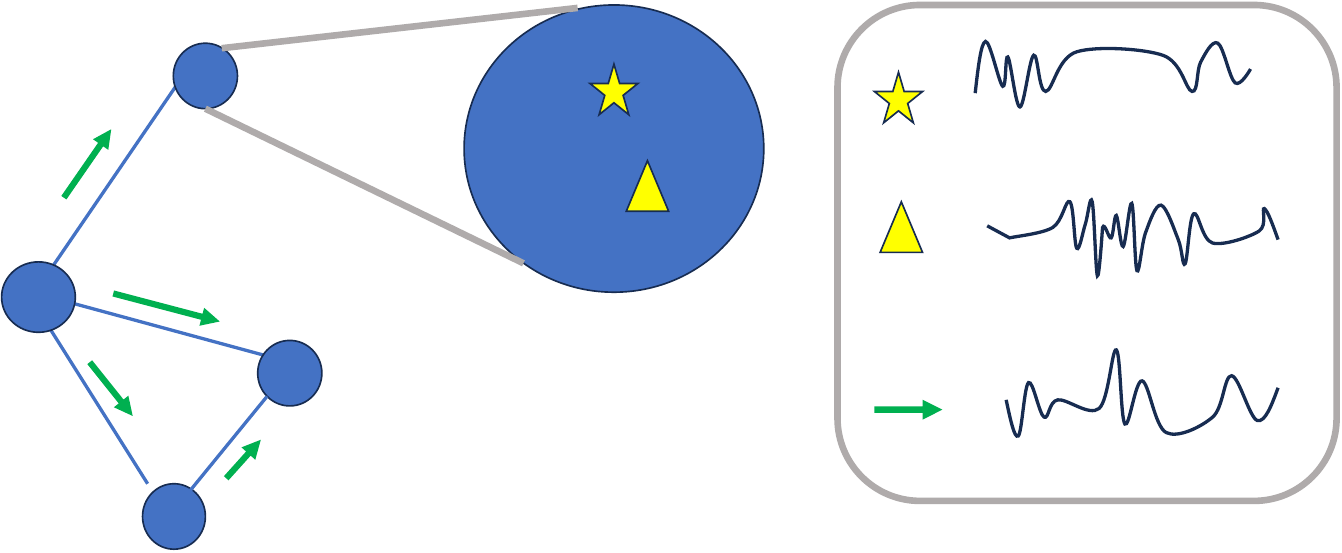}
\caption{\textbf{Schematic representation of the dynamical state of the network for dynamical Turing patterns.} We consider a network $G=(V,E)$ whose dynamical states includes  two node topological signals (species densities represented with yellow triangles and yellow stars respectively on the enlarged node, although they are defined on each node) and one link topological signal  schematically represent with  green arrows associated to links.}
\label{fig:scheme}
\end{figure}

\section{The 3-way Dirac operator: 2 species on the nodes and 1 species on the links}\label{sec:2nod1link}

 In this work we consider a network dynamical state captured by two topological signals on the nodes and one topological signal on the links encoded by the topological spinor ${\Phi}\in C^{0} \oplus C^{0} \oplus C^{1}$ (see Fig.\ref{fig:scheme}). Note that this approach can be readily generalized to the case in which the dynamical state of the network is encoded into a single topological node signal and two topological link signals. 

The topological spinor  encodes the 3-way topological signals of the network and has block structure
\bea
{\mathbf{\Phi}}=\left(\begin{array}{c}
\bm \chi \\
\bm \psi
\end{array}\right)\, ,
\eea
where $\bm \chi\in C^0\oplus C^0$ is defined on nodes and $\bm \psi\in C^1$ is defined on links. 
In particular we will use the notation
\bea
\bm \chi=
\left(\begin{array}{c}
{u} \\
{v}
\end{array}\right),\quad \bm \psi={w}\, ,
\eea
with $u\in C^0$, $v\in C^0$ indicating the two node signals and $w\in C^1$ indicating the link signal.
For ease of notation here and in the following we indicate with $\mathcal{N}=2N_{0}+N_{1}$ and with $\mathcal{M}=N_0+2N_1$.
We have therefore that $\bm \Phi$ is the $\mathcal{N}$ dimensional column vector, given by
\bea
\bm \Phi= \left(\begin{array}{c}
{u} \\
{v}\\
{w}
\end{array}\right)\, .
\eea
The 3-way Hodge-Dirac operator $\bm\partial:C^0\oplus C^0\oplus C^1\to C^0\oplus C^{1}\oplus C^1$ acts on the topological spinor $\Phi$ projecting  independently the two node signals encoded by $\bm \chi$ into the links and the single link signal $\bm \psi$ into the nodes.
We define the 3-way Hodge-Dirac operator $\bm \partial$ as the $\mathcal{M}\times \mathcal{N}$ matrix given by  
\bea
\bm{\partial}=\left(\begin{array}{ccc}
{\bf 0} &  {\bf I}_1\otimes\mathbf{B} \\
{\bf I}_2\otimes\mathbf{B}^{\top} & {\bf 0}\end{array}\right)\, ,
\eea
or alternatively as
\bea
\bm{\partial}=\left(\begin{array}{ccc}
0 & 0 & \mathbf{B} \\
\mathbf{B}^{\top}&0 & 0\\
0&\mathbf{B}^{\top} &   0
\end{array}\right)\, .
\eea
In contrast with the Hodge-Dirac operator, the  Dirac operator ${\Dirac}:C^0\oplus C^0\oplus C^1\to C^0\oplus C^0\oplus C^1$ allows the node and link signals to cross talk as it maps topological spinors into topological spinors. The Dirac operator is obtained by multiplying the Hodge-Dirac operator by the $\mathcal{N}\times \mathcal{M}$ gamma matrix $\bm \gamma$, i.e. 
\bea
\Dirac=\bm\gamma{\bm{\partial}}\, .
\eea
Therefore the gamma matrix $\bm \gamma$ 
defines the way in which the two nodes signals projected on the links are compressed and how the single link signal projected on the nodes is expanded into the two nodes signals of the topological spinor.
The gamma matrix $\bm \gamma$ is defined in terms of 2-dimensional column vector ${\bm \alpha}=(\alpha_u,\alpha_v)^{\top}$ and a 2-dimensional row vector ${\bm\beta}=(\beta_u,\beta_v)$, as
\bea
\bm\gamma=\left(\begin{array}{ccc}
{\bm \alpha}\otimes {\bf I}_{N_0} & {\bf 0} \\
{\bf 0} & {\bm\beta} \otimes {\bf I}_{N_1}
\end{array}\right)\, ,
\eea
or alternatively
\bea
\bm\gamma=\left(\begin{array}{ccc}
{\alpha}_u {\bf I}_{N_0} & 0 &0\\
{\alpha}_v {\bf I}_{N_0} & 0 &0\\
{0} & {\beta}_u  {\bf I}_{N_1} &{\beta}_v  {\bf I}_{N_1}
\end{array}\right)\, .
\eea
It follows that the 3-way Dirac operator can be also expressed as 
\begin{equation}
\Dirac=\bm\gamma\bm\partial=\left(\begin{array}{ccc}
0 & 0 & \alpha_u\mathbf{B} \\
0 & 0 & \alpha_v\mathbf{B} \\
\beta_u\mathbf{B}^{\top}&\beta_v\mathbf{B}^{\top} & 0
\end{array}\right)\, .
\end{equation}

As previously shown in Eq.~\eqref{eq:delta2L} the  Dirac operator defined on a topological spinors having the same number of topological signals on nodes and links, can be interpreted as the ``square-root'' of the higher-order Laplacian. Therefore it is natural to investigate whether a similar condition holds true for 3-way Dirac operator. One can straightforwardly obtain
\begin{equation}
\Dirac^{2}=\bm{\mathcal{L}}=\left(\begin{array}{ccc}
D_{uu}\mathbf{L}_{[0]} & D_{uv}\mathbf{L}_{[0]} & 0 \\
D_{vu}\mathbf{L}_{[0]} & D_{vv}\mathbf{L}_{[0]} & 0 \\
0 & 0 & D_{ww} \mathbf{L}_{[1]}
\end{array}\right)\, ,
\end{equation}
with $D_{uu}=\alpha_u\beta_u,D_{uv}=\alpha_u\beta_v,D_{vu}=\alpha_{v}\beta_u,D_{vv}=\alpha_v\beta_v$ and $D_{ww}=\alpha_u\beta_u+\alpha_v\beta_v$, assumed to be non-negative.
It is instructive to consider the case in which $\bm\alpha=(1,1)^{\top}$ and $\bm\beta=(1,1)$.
In this case we have  $\Dirac=\Dirac_{\ 0}$ with 
\bea
\Dirac_{\ 0}=\left(\begin{array}{ccc}
0 & 0 & \mathbf{B} \\
0 & 0 & \mathbf{B} \\
\mathbf{B}^{\top}&\mathbf{B}^{\top} & 0
\end{array}\right)\text{ and  } \Dirac_{\ 0}^2=\bm{\mathcal{L}}_0=\left(\begin{array}{ccc}
\mathbf{L}_{[0]} & \mathbf{L}_{[0]} & 0 \\
\mathbf{L}_{[0]} & \mathbf{L}_{[0]} & 0 \\
0 & 0 & 2 \mathbf{L}_{[1]}
\end{array}\right).
\label{D0}
\eea

On a side note we observe  that the Dirac operator defined in Eq.(\ref{D0}) is reminiscent of the   topological operator used  for studying duplex networks  with link overlap in Ref. \cite{ghorbanchian2022}.

Given the diagonal block structure of $\bm{\mathcal{L}}$, its eigenvalues are either the eigenvalues of $\bm{\mathcal{L}}_{uv}$ or the ones of $\bm{\mathcal{L}}_{w}$ with 
\bea
\bm{\mathcal{L}}_{uv}=\left(\begin{array}{cc}
D_{uu}\mathbf{L}_{[0]} & D_{uv}\mathbf{L}_{[0]}  \\
D_{vu}\mathbf{L}_{[0]} & D_{vv}\mathbf{L}_{[0]} 
\end{array}\right),\quad \bm{\mathcal{L}}_{w}=D_{ww} \mathbf{L}_{[1]}\, .
\eea

From this property it follows that the eigenvalues of the Dirac operator are the square roots of the eigenvalues of $\mathcal{L}$ taken both with positive and negative sign. Therefore, while the Hodge Laplacians are semi-definite positive the Dirac operator is not semi-definite positive and, as we will discuss in the following, might also display complex eigenvalues.

We observe that the three way Dirac operator $\Dirac~$ commutes with the matrix $\bm\gamma_0$ given by
\begin{equation}
\bm\gamma_0=\left(\begin{array}{ccc}
\mathbf{I}_{N_0} & 0 & 0 \\
0 & \mathbf{I}_{N_0} & 0 \\
0 & 0 & -\mathbf{I}_{N_1}
\end{array}\right)\, .
\end{equation}

Hence if a spinor $\boldsymbol{\Phi}=(u, v, w)^{\top}$ is an eigenvector of the three way Dirac operator with eigenvalue ${\hat\lambda} \neq 0$, i.e. if
\begin{equation}
\Dirac \bm \Phi={\hat\lambda} \bm\Phi\, ,
\end{equation}
then $\bm\gamma_0 \mathbf{\Phi}=(u, v,-w)^{\top}$ is an eigenvector of the three way Dirac operator with eigenvalue $-{\hat\lambda}$. Indeed we have
\begin{equation}
\Dirac \bm\gamma_0 \bm \Phi=-\bm\gamma_0 \Dirac \bm \Phi=-{\hat\lambda} \bm\gamma_0\bm \Phi\, .
\end{equation}

From these results it follows that the harmonic eigenvectors of the three way Dirac operator are
\begin{equation}
\phi_{0}^{\text {harm }}=\left(\begin{array}{c}
\bm{\chi}^{\text{harm}} \\
0
\end{array}\right) \quad \phi_{1}^{\text {harm }}=\left(\begin{array}{c}
0 \\
{\bm \psi}^{\text {harm }} \\
\end{array}\right)\, ,
\end{equation}
where $\bm \chi^{\text{harm}}$ indicates the generic harmonic eigenvector of $\bm{\mathcal{L}}_{uv}$ while ${\bm \psi}^{\text {harm }}$ the generic harmonic eigenvector of $\bm{\mathcal{L}}_{w}$.

Note that the harmonic eigenvectors $\bm\chi^{\text{harm}}$ read
\bea
\left(\begin{array}{c}
\hat{\bm \chi}_0 \\
0 \\
\end{array}\right),\quad \left(\begin{array}{c}
0 \\
\hat{\bf \chi}_0 
\end{array}\right),\quad \left(\begin{array}{c}
-\beta_v\hat{\bm\chi} \\
\beta_u \hat{\bm \chi} 
\end{array}\right)\, . 
\eea
Where $\hat{\bm\chi}_0$ are the harmonic right singular vectors of the boundary matrix, and where $\hat{\bm\chi}$ is an arbitrary $N_0$ dimensional vector.
The non zero eigenvalues $\hat\lambda_k$ of the $3$-way Dirac operator  are given by  
 \bea
 \hat\lambda_k=\pm\sqrt{\alpha_u\beta_u+\alpha_v\beta_v}\mu_k
 \eea
 where $\mu_k$ are the singular eigenvalues of the boundary operator.
Note that with a general choice of the gamma matrix $\bm\gamma$ the eigenvalues of the Dirac operator are real only if 
\bea
\Delta=\alpha_u\beta_u+\alpha_v\beta_v>0,
\eea
and becomes purely imaginary (i.e., appears in complex conjugate pairs) for $\Delta<0$ while they are identically zero in the trivial case $\Delta=0$.

The eigenvectors $\boldsymbol{\phi}^{+}$, $\boldsymbol{\phi}^{-}$ associated to the non-zero eigenvalues $\hat\lambda_k$ with the same absolute value are related by chirality that for the three-way Dirac operator takes the form
\begin{equation}
\phi^{+}=\mathcal{N}\left(\begin{array}{c}
\alpha_u\hat{\bm \chi}_{\mu} \\
\alpha_v\hat{\bm \chi}_{\mu}\\
\sqrt{\Delta} {\bm \psi}_{\mu}
\end{array}\right) \quad \phi^{-}=\mathcal{N}\left(\begin{array}{c}
\alpha_u\hat{\bm \chi}_{\mu} \\
\alpha_v\hat{\bm \chi}_{\mu}\\
-\sqrt{\Delta} {\bm \psi}_{\mu}
\end{array}\right)\, , 
\end{equation}
where $\hat{\chi}_{\mu}$ are the right singular vectors and $\bm\psi_{\mu}$ are the left singular vectors corresponding to a non-zero singular value $\mu=\mu_k$ of the boundary operator.

Let us now discuss the spectral properties of the Dirac operator $\Dirac=\Dirac_{\ 0}$ defined in Eq. (\ref{D0}) in the case in which $\bm\alpha=(1,1)^{\top}$ and $\bm\beta=(1,1)$. As long as the network is  connected  the zero eigenvalue of ${\bm{\mathcal{L}}}$ has multiplicity $N_{1}+2$ while the non-zero eigenvalues  are given by twice the non-zero eigenvalues $\mu$ of $L_{[0]}$ (or equivalently of $\mathbf{L}_{[1]}$ since on a network $\mathbf{L}_{[0]}$ is isospectral to $\mathbf{L}_{[1]}$). Therefore the eigenvalues of the Dirac operator $\Dirac_{\ 0}$ in this case are real and given by 
\begin{equation}
\begin{aligned}
& \hat\lambda_k=0 \quad \text { with multiplicity } N_{1}+2, \\
& \hat\lambda_k= \pm \sqrt{2}\mu_k\, .
\end{aligned}
\end{equation}

\section{Theory of topological 3-way Turing patterns}
\label{ssec:2no1link}
Our goal will be now to define a dynamical system for the $3$-way topological spinor defined in the previous section, that can display dynamical Turing patterns on nodes and links.
To this end, we assume that in absence of interactions, the dynamical state of the network, captured by the topological spinor $\bm\Phi$ follows the dynamics 
\bea\label{eq:syst_2nod1link_reaction}
 \dot{\bm \Phi}={\bf F}(\bm\Phi,\bm\partial\bm\Phi),
\eea
where  ${\bf F}(\bm\Phi,\bm\partial \bm\Phi)$ determines the reaction dynamics  involving the three signals $u,v,w$. Let us observe that, as we will see further in this section, the term $\bm\partial \bm\Phi$ vanishes when there are no interactions between simplices of different dimensions. Nonetheless, this formalism allows us to consider also nonlinear terms in $\mathbf{B}$ and $\mathbf{B}^\top$, making the latter more general. Let us now specify more in detail the assumed structure of this reaction term.
As previously stated the topological spinor $\bm \Phi=(\bm \chi,\bm \psi)^{\top}$, where $\bm\chi=(u,v)^{\top}$, encodes for the two node signals and $\bm\psi=w$ encodes for the single link signal.
On each node the signal $u$ and $v$ can interact, while they cannot interact with the link signal.
The projected topological spinor $\bm\partial \bm\Phi$ is instead given by 
\bea
\bm\partial \bm\Phi=\left(\begin{array}{c}\bm \hat{\bm\chi}\\\hat{\bm\psi}\end{array}\right),\eea
with 
\bea
\hat{\bm\chi}={\bf B}w, \quad \hat{\bm\psi}=\left(\begin{array}{c}{\bf B}^{\top}u\\{\bf B}^{\top}v\end{array}\right).
\eea
In other words $\hat{\bm\chi}$ is the one dimensional projected signal on the nodes and $\hat{\bm\psi}$ is the two dimensional projected signals on the links.
Since on nodes only node signals can interact, and on links only link signals can interact,
it follows that on nodes all the signals encoded in $\bm\chi$ can interact with each other and with the projected signals $\hat{\bm\chi}$ and similarly on the links the projected signals on the links $\hat{\bm\psi}$ can interact with each other and with the link signal ${\bm\psi}$.

Accordingly, our choice of the reaction term  ${\bf F}(\bm\Phi,\bm \partial \bm\Phi)$ is
\bea
{\bf F}(\bm\Phi,\bm \partial \bm\Phi)=\left(\begin{array}{c}f_1({\bm\chi},\hat{\bm\chi})\\f_2({\bm\chi},\hat{\bm\chi})\\f_3({\bm\psi},\hat{\bm\psi})\end{array}\right)=\left(\begin{array}{c}f(u,v,{\bf B}w)\\g(u,v,{\bf B}w)\\h({\bf B}^{\top}u,{\bf B}^{\top}v,w)\end{array}\right),\eea 
where $f,g,h$ act on their arguments element-wise.

Here we want to show that topological dynamical Turing patterns emerge when we introduce interactions between the topological signals of nodes and nearby links  captured by the Laplacian and Dirac operators. We therefore consider the reaction-diffusion equations

\begin{equation}\label{eq:syst_2nod1link}
    \dot{\bm \Phi}={\bf F}(\bm\Phi,\bm \partial \bm\Phi)-c_1\Dirac \bm\Phi-c_2\bm{\mathcal{L}}\bm\Phi,
\end{equation}
with  $c_1\in \mathbb{R}$, $c_2\in \mathbb{R}^+$.
Note that when $c_1=0$ the interactions will be only diffusive, i.e., driven by the Hodge Laplacians while when $c_2=0$ the interaction is  exclusively driven by the Dirac operator. The possible patterns emerging in the former case will thus named Turing patterns, while we will deal with Dirac patterns, in the latter scenario. Let us observe that interestingly, this latter scenarios that we introduce here, leads to very non trivial dynamical patterns as we will show in the next sections. Note that these Dirac induced patterns  have no equivalent in the framework of node centered theory of Turing pattern formation.

We can rewrite Eq.~\eqref{eq:syst_2nod1link} explicitly, obtaining the following $3$-species reaction-diffusion system 
\begin{equation}\label{eq:rd_2nod1link}
    \begin{cases}
      \dot{u}=f(u,v,\mathbf{B}w)-c_2\mathbf{L}_{[0]}(D_{uu} u+ D_{uv} v)-c_1\alpha_{u}\mathbf{B}w, \\
      \dot{v}=g(u,v,\mathbf{B}w)-c_2\mathbf{L}_{[0]}(D_{vu} u + D_{vv} v)-c_1\alpha_{v}\mathbf{B}w , \\
      \dot{w}=h(\mathbf{B}^\top u,\mathbf{B}^\top  v,w)-c_2D_{ww}\mathbf{L}_{[1]}-c_1\mathbf{B}^{\top}(\beta_u u+\beta_v v)\, .
    \end{cases}
\end{equation}
Let us observe that cross-diffusion emerges naturally from the formalism developed in the previous section. \\

\paragraph*{Remark} Let us observe that, if we take functions $f$, $g$ and $h$ such that the terms in $\mathbf{B}$ and $\mathbf{B}^\top$ are linear (as it will be the case for the example that we will hereafter present), we can map the problem into a new one with a local reaction term ${\bf F}(\bm\Phi)$ and a linear Dirac part. The relevance of this interpretation will become clear in the next Sections, when we will show that an oscillatory instability can emerge solely due to the Dirac coupling. \\

In order to study the emergence of Turing and Dirac patterns, we need the system to exhibit a stable homogeneous equilibrium and that turns unstable once subjected to a spatially inhomogeneous perturbation. Before studying the stability of the homogeneous state, let us recall that such equilibrium is a solution for the interconnected system provided the constant eigenvector ${\bf 1}$ is in the kernel of the Dirac operator \cite{turing_topological,carletti2023global}. For the case of networks, this implies that the divergence of the constant link signal is null, i.e., that there is an orientation of the simplex such that for each node $r$ a equal number of links  point toward it and point outward it \cite{turing_topological}.  This relation implies that the network must be Eulerian, i.e., to have only nodes with even degree.

 When the coupling is not active, we have $\bm\partial \bm\Phi^{\star}=(0,0,0)$, hence, the system reduces to $N_0$ isolated systems of two interacting species on the nodes and $N_1$ systems of one species on the links, of the following form 
\begin{equation}\label{eq:2nod1link_isolated}
    \begin{cases}
      \dot{u}=f(u,v,0), \\
      \dot{v}=g(u,v,0), \\
      \dot{w}=h(0,0,w)\, ,
    \end{cases}
\end{equation} 
hence, we can study the local stability of this homogeneous state considering nodes and links separately. \\ Let us assume $\bm\Phi^*=(u^*,v^*,w^*)^\top$ to be a fixed point for the above system, meaning that $w^*$ is a fixed point of $h(0,0,w)$ and $(u^*,v^*)^\top$ for the system $f(u,v,0),~g(u,v,0)$. For the equation on the links, the linear stability condition is trivially $h_w<0$, where with $h_w$ we indicate the derivative of function $h$ with respect to variable $w$ evaluated at the homogeneous state $w^*$. For what concerns the dynamics on the nodes, it reduces to studying the stability of the following $2$ dimensional system 
\begin{displaymath}
     \begin{cases}
      \dot{u}=f(u,v,0) \\
      \dot{v}=g(u,v,0)\, . 
      \end{cases}
\end{displaymath}
From the assumption that $(u^*,v^*)^\top$ is a fixed point for the above system, we can linearize around it and obtain the Jacobian matrix, whose stability conditions are $f_u+g_v<0$ and $f_ug_v-f_vg_u>0$. Hence, the homogeneous vector $(u^*,v^*,w^*)^\top$ is a stable equilibrium for system \eqref{eq:2nod1link_isolated} provided that \begin{equation}\label{eq:hom_eq}
    \begin{cases}
f_u+g_v<0\\
f_ug_v-f_vg_u>0\\
h_w<0\ .
        \end{cases}
\end{equation}

Let us now perturb system \eqref{eq:rd_2nod1link} with a spatially inhomogeneous perturbation $(\delta u,\delta v, \delta w)^\top$ and linearize it around the homogeneous equilibrium point that we have proved to be a solution also of the coupled system. We hence obtain the following system for the dynamics of the perturbation  
\begin{equation}\label{eq:rd_2nod1link_lin}
    \begin{cases}
      \delta\dot{u}=f_u \delta u + f_v \delta v +f_w\mathbf{B}\delta w-c_2 \mathbf{L}_{[0]}(D_{uu}\delta u+ D_{uv}\delta v) \\
      \delta\dot{v}=g_u \delta u + g_v \delta v +g_w\mathbf{B}\delta w-c_2\mathbf{L}_{[0]}(D_{vu}\delta u+ D_{vv}\delta v) \\
      \delta\dot{w}=h_{u}\mathbf{B}^\top\delta u+h_{v}\mathbf{B}^\top\delta v +h_w \delta w-c_2D_{ww}\mathbf{L}_{[1]}\delta w\, ,
    \end{cases}
\end{equation}
where, for sake of notation, we have considered the terms in $c_1$ of Eq. \eqref{eq:rd_2nod1link} as part of the reaction functions, namely $f_w=\partial_{\mathbf{B}w} f -c_1\alpha_u$, $g_w=\partial_{\mathbf{B}w}g-c_1\alpha_v $, $h_{u}=\partial_{\mathbf{B}^\top u}h-c_1\beta_u$ and $h_{v}=\partial_{\mathbf{B}^\top v}h-c_1\beta_v$.  \\

We can proceed as in \cite{turing_topological} and use the eigenvectors of $\mathbf{L}_{[0]}$ and $\mathbf{L}_{[1]}$ to perform the singular value decomposition of $\mathbf{B}$. Let us recall that $\mathbf{L}_{[0]}$ and $\mathbf{L}_{[1]}$ are, in this case, isospectral and their eigenvalues $\mu_k^2$ are the square of the singular values $\mu_k$ of $\mathbf{B}$. By projecting the perturbations $\delta u$ and $\delta v$ on the eigenbasis of $\mathbf{L}_{[0]}$, and the perturbation $\delta w$ on the eigenbasis of $\mathbf{L}_{[1]}$, Eq. \eqref{eq:rd_2nod1link_lin} becomes 
\begin{equation}
\label{eq:rd_2nod1link_lin_goodbasis}
    \begin{cases}
      \delta\dot{\hat{u}}_k=f_u \delta \hat{u}_k + f_v \delta \hat{v}_k +f_w \mu_k\delta \hat{w}_k-\mu_k^2c_2 (D_{uu}\delta \hat{u}_k+ D_{uv}\delta \hat{v}_k), \\
      \delta\dot{\hat{v}}_k=g_u \delta \hat{u}_k + g_v \delta \hat{v}_k +g_w \mu_k\delta \hat{w}_k-\mu_k^2c_2(D_{vu}\delta \hat{u}_k+ D_{vv}\delta \hat{v}_k), \\
      \delta\dot{\hat{w}}_k=h_{u} \mu_k\delta \hat{u}_k+h_{v} \mu_k\delta \hat{v}_k +h_w \delta \hat{w}_k-\mu_k^2c_2D_{ww}\delta \hat{w}_k,
    \end{cases}
\end{equation} where $\delta\hat{u}_k$ (resp. $\delta\hat{v}_k,\delta\hat{w}_k$) is the perturbation expressed in the new basis. 
To study the linear stability of the perturbation, we hypothesize that $\delta \hat{u}_k(t),\delta \hat{v}_k(t),\delta \hat{w}_k(t)\sim e^{\lambda_k t}$. To ensure the existence of a nontrivial solution, the linear growth rate $\lambda_k$ must satisfy the following condition
\begin{equation}\label{eq:nontriv}
    \left|J_k\right|=0\, ,
\end{equation}
where $|J_k|$ denotes the determinant of the matrix
\begin{displaymath}
   J_k=\left(\begin{array}{ccc}
f_u-\mu_k^2c_2D_{uu}-\lambda_k & f_v-\mu_k^2c_2D_{uv} & \mu_k f_w \\ g_u-\mu_k^2c_2D_{vu} & g_v-\mu_k^2c_2D_{vv}-\lambda_k & \mu_k g_w \\ \mu_kh_u & \mu_kh_v & h_w-\mu_k^2c_2D_{ww}-\lambda_k
\end{array}\right)\, .
\end{displaymath} 
We can rewrite this condition as a $3$-rd order polynomial in the variable $\lambda_k$ of the form 
\begin{equation}\label{eq:poly}
    a \lambda_k^3 + b\lambda_k^2 + c\lambda_k +d = 0,
\end{equation}
whose coefficients are
\begin{displaymath}
\begin{cases}
      a=1 \\
    b=c_2\mu_k^2(D_{uu}+D_{vv}+D_{ww})-(f_{u}+g_{v}+h_{w})\\
    c= A \mu_k^4+ B \mu_k^2 +C \equiv \pi_1(\mu_k^2)\\
    d= \mathcal{A}\mu_k^4 + \mathcal{B}\mu_k^2 +\mathcal{C} \equiv \pi_2(\mu_k^2)\, ,
    \end{cases}
\end{displaymath}
where $\pi_1$ and $\pi_2$ are polynomials in $\mu_k^2$, whose explicit expressions can be found in \ref{app:A}. By means of the Routh-Hurwitz criterion \cite{Routh1877,Hurwitz1895}, we can study the stability of polynomial \eqref{eq:poly}. When it is unstable, we have Turing patterns. The explicit conditions are rather cumbersome, but it is easy to check that Turing patterns can be obtained.\\

We will call $\lambda$ the maximum real part of the $\lambda_k$, and its associate imaginary part $\varrho$. Here $\lambda$ and $\varrho$ are the real and imaginary part of the dispersion relation, respectively, and they are both functions of the continuous parameter $\mu_k$. The condition for having Turing instability is thus the existence of a $\mu_k$ for which $\lambda(\mu_k)>0$. Let us observe that being the support discrete, there can be finite size effects, namely $\lambda(\mu_k)<0$ for all $k$ while the continuous function $\lambda$ can assume positive values. This means that the networked system may not exhibit Turing patterns, while the same exact system defined on a continuous support will develop patterns. 

Moreover, if $\lambda(\mu_k)>0$ and the corresponding imaginary part is non-zero, i.e., $\varrho(\mu_k)\neq 0$, then the system can exhibit dynamical (oscillatory) patterns~\cite{hata_wave}. Note that since we are considering case for which $J_k$ is a real valued matrix, any non-zero value of $\varrho$ implies also the presence of a solution with $-\varrho$. Let us finally remark that a non-zero imaginary part of a critical mode is not a sufficient condition for the existence of wave patterns, as there are cases in which the patterns are stationary despite a non-zero $\varrho$ \cite{jop_carletti}. In fact, one cannot know \textit{a priori} which kind of pattern is obtained by solely using the information contained in the linear stability analysis, and indeed the problem of pattern prediction is still open.

\section{Emergence of dynamical Turing patterns on nodes and links}
\label{ssec:2no1link_numerical}

In order to show evidence of the emergence of dynamical topological Turing pattern, we consider the following dynamical system inspired by the excitable dynamics of the FitzHugh-Nagumo neuronal model \cite{fitzhugh,nagumo}  
\begin{equation}\label{eq:rd_2nod1link_model}
    \begin{cases}
      \dot{u}=\sigma_1 u-\eta_1 u^3+\xi_1v+\zeta_1\mathbf{B}w-c_2\mathbf{L}_{[0]}(D_{uu} u+ D_{uv} v)-c_1\alpha_{u}\mathbf{B}w \\
      \dot{v}=\sigma_2v+\xi_2 u+\zeta_2 \mathbf{B}w-c_2\mathbf{L}_{[0]}(D_{vu} u + D_{vv} v)-c_1\alpha_{v}\mathbf{B}w  \\
      \dot{w}=\sigma_3  w+\zeta_3\mathbf{B}^{\top}u+\zeta_4\mathbf{B}^{\top}v-c_2D_{ww}\mathbf{L}_{[1]}-c_1\mathbf{B}^{\top}(\beta_u u+\beta_v v)\, ,
    \end{cases}
\end{equation}
whose homogeneous equilibrium state is given by \begin{displaymath}
    (u^*,v^*,w^*)=(0,0,0)\, .
\end{displaymath}

\begin{remark}
Let us observe that in this particular case, the homogeneous vector $(0,0,0) \otimes (\mathbf{1}_{N_0},\mathbf{1}_{N_0},\mathbf{1}_{N_1})$ trivially belongs to the kernel of the Dirac operator for any network. However, for sake of continuity with the general framework above presented, we decided to apply the model~\eqref{eq:rd_2nod1link_model} to an Eulerian graph as support.

Let us observe that a similar claim could be applied any time the homogeneous equilibrium is of the form $(u^*,v^*,0)$, because the network Laplace matrix $\mathbf{L}_{[0]}$ admits $(\mathbf{1}_{N_0},\mathbf{1}_{N_0})$ as eigenvector associated to the $0$ eigenvalue for any connected network.
\end{remark}

In order for the homogeneous equilibrium on the nodes to exist, we need the network to be connected. The same condition for the variables on the links is attained if the network is Eulerian, i.e., every node has an even degree, as shown in \cite{turing_topological}. Let us observe that the equilibrium of the chosen model would allow to relax the latter condition, nonetheless we want our theory to be general.
The support on which the dynamics take place is a $2$D lattice with periodic boundary condition of $36$ nodes, i.e., $6\times 6$. \\

The conditions~\eqref{eq:hom_eq} ensuring the stability of the homogeneous equilibrium become in the present case
\begin{equation}
\label{eq:bif_hom}
    \begin{cases}
        \sigma_2<-\sigma_1\\
        \displaystyle\sigma_2>\frac{\xi_2\xi_1}{\sigma_1}\\
        \sigma_3<0\, .
    \end{cases}
\end{equation}
As shown in \ref{app:A}, although obtaining the  explicit form of the conditions for Turing instability is challenging, relying on numerical results is possible.

Here we show results of dynamical Turing and Dirac pattern on two different Eulerian network structures: a $2D$ square lattice tessellating a torus and random graph with given degree sequence.

\subsection{Phenomenology on a torus (2D square lattice with periodic boundary conditions)}

\begin{figure}[ht!]
\centering
\includegraphics[scale=1]{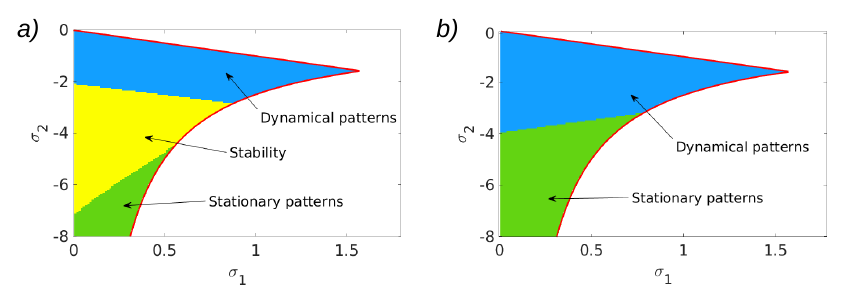}
\caption{\textbf{Instability regions in the $(\sigma_1,\sigma_2)$ space} The red curves are the bifurcation curves determining the stability of the homogeneous equilibrium, namely, the first two equations of system \eqref{eq:bif_hom}. Panel a) shows the case with diffusion (Turing patterns), while the case of Dirac-induced patterns is shown in panel b). Having fixed $\sigma_3<0$, the homogeneous equilibrium is unstable in the white region, while it is stable in the colored area. The blue shaded area is where the patterns are oscillatory, while in the green one they are stationary. The yellow region is where the homogeneous equilibrium remains stable even after the spatially inhomogeneous perturbation, which never occurs for Dirac-induced patterns with this set of parameters. The colored regions are obtained by computing the real dispersion relation and, if it is positive, checking if the imaginary part of the positive modes is zero (green region) or nonzero (blue region). The parameters are $\eta_1=-3,~\xi_1=-0.5,~\xi_2=5,
~\zeta_1=0.1,~\zeta_2=6,~\zeta_3=6,~\zeta_4=1,~c_1=-1,~\alpha_u=1,~\alpha_v=1,~\beta_u=1,~\beta_v=0.1$; for panel a) we have $\sigma_3=-2,~c_2=1$, while for panel b) $\sigma_3=-10,~c_2=0$.}
\label{fig:Tur_Dir_regions}
\end{figure}

We consider a $2D$ square lattice tessellating the torus. Specifically we assume to deal with a $6\times 6$ square lattice with periodic boundary conditions.
 In Fig. \ref{fig:Tur_Dir_regions}, we report the bifurcation diagram of the model in the parameter space $(\sigma_1,\sigma_2)$  
 computed from the dispersion relation for the cases in which both $c_1$ and $c_1$ are non-zero (panel a) and 
 in which the interaction is exclusively driven by the Dirac operator, i.e., $c_2=0$ (panel b).

In both bifurcation diagrams the red curves, i.e., $\sigma_1+\sigma_2=0$ and $\sigma_1\sigma_2 =\xi_1\xi_2$, determine the boundary of the stability region of the homogeneous equilibrium, which is unstable outside the region (white region). The system exhibits patterns for parameters in the green and blue regions. In the latter, patterns are dynamics, while in the former they are stationary, because $\varrho$, i.e., the imaginary part associated to unstable mode $\lambda$, is zero. The yellow region determines parameters values for which the homogeneous equilibrium remains stable, even after an inhomogeneous perturbation; hence, no patterns are observed.  Note that, for this choice of parameters, the latter region is observed in presence of diffusion, i.e., $c_2>0$, while it disappears in the case of the exclusive Dirac coupling, i.e. $c_2=0$ (see panel b)).

\begin{figure}[ht!]
\centering
\includegraphics[scale=1]{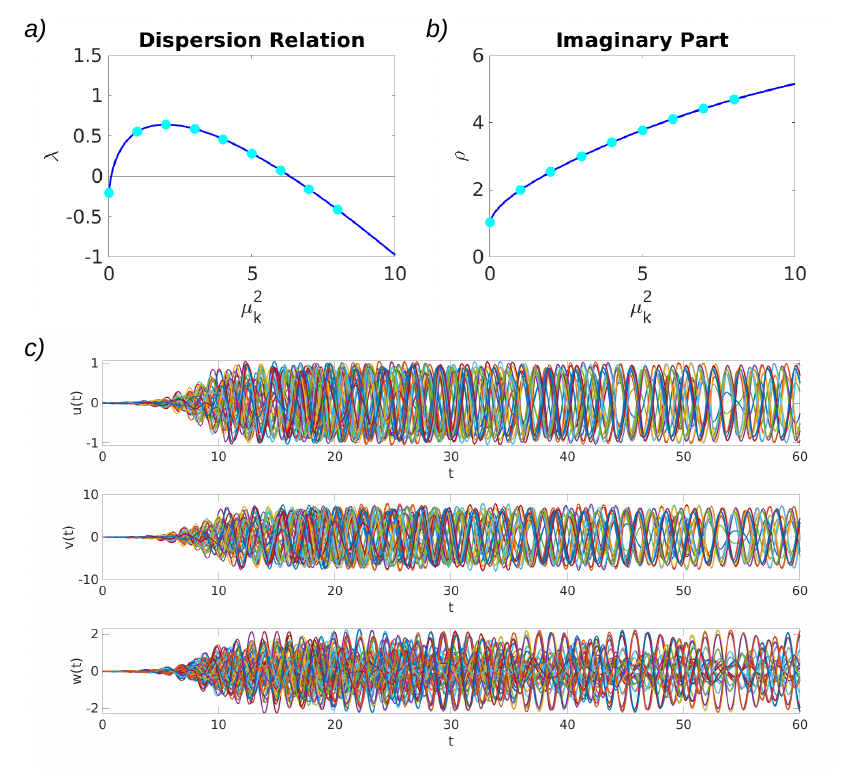}
\caption{\textbf{Dispersion relation and dynamical Turing patterns} Panel a) shows the real part of the dispersion relation $\lambda$, while panel b) its respective imaginary part $\varrho$. The blue curves are computed by considering a continuous parameter $\mu_k$, while the cyan dots indicate the eigenvalues of the Hodge Laplacians. The latter determine the effective dispersion relation and the emergence or not of Turing patterns. The chosen setting is such that Turing patterns emerge ($\lambda>0$) and they are oscillatory (the corresponding $\varrho\neq 0$), as shown by the dynamics of the $3$ interacting species, in panel c). The network is a $2$D lattice with periodic boundary conditions of $36$ nodes and the parameters are $\sigma_1=1,~\sigma_2=-1.4,~\sigma_3=-2,~\eta_1=-3,~\xi_1=-0.5,~\xi_2=5,
~\zeta_1=0.1,~\zeta_2=6,~\zeta_3=6,~\zeta_4=1,~c_1=-1,~c_2=1,~\alpha_u=1,~\alpha_v=1,~\beta_u=1,~\beta_v=0.1$; the initial perturbation is the order of $10^{-2}$.}
\label{fig:rel_disp_time_series_Tur}
\end{figure}

\begin{figure}[ht!]
\centering
\includegraphics[scale=0.3]{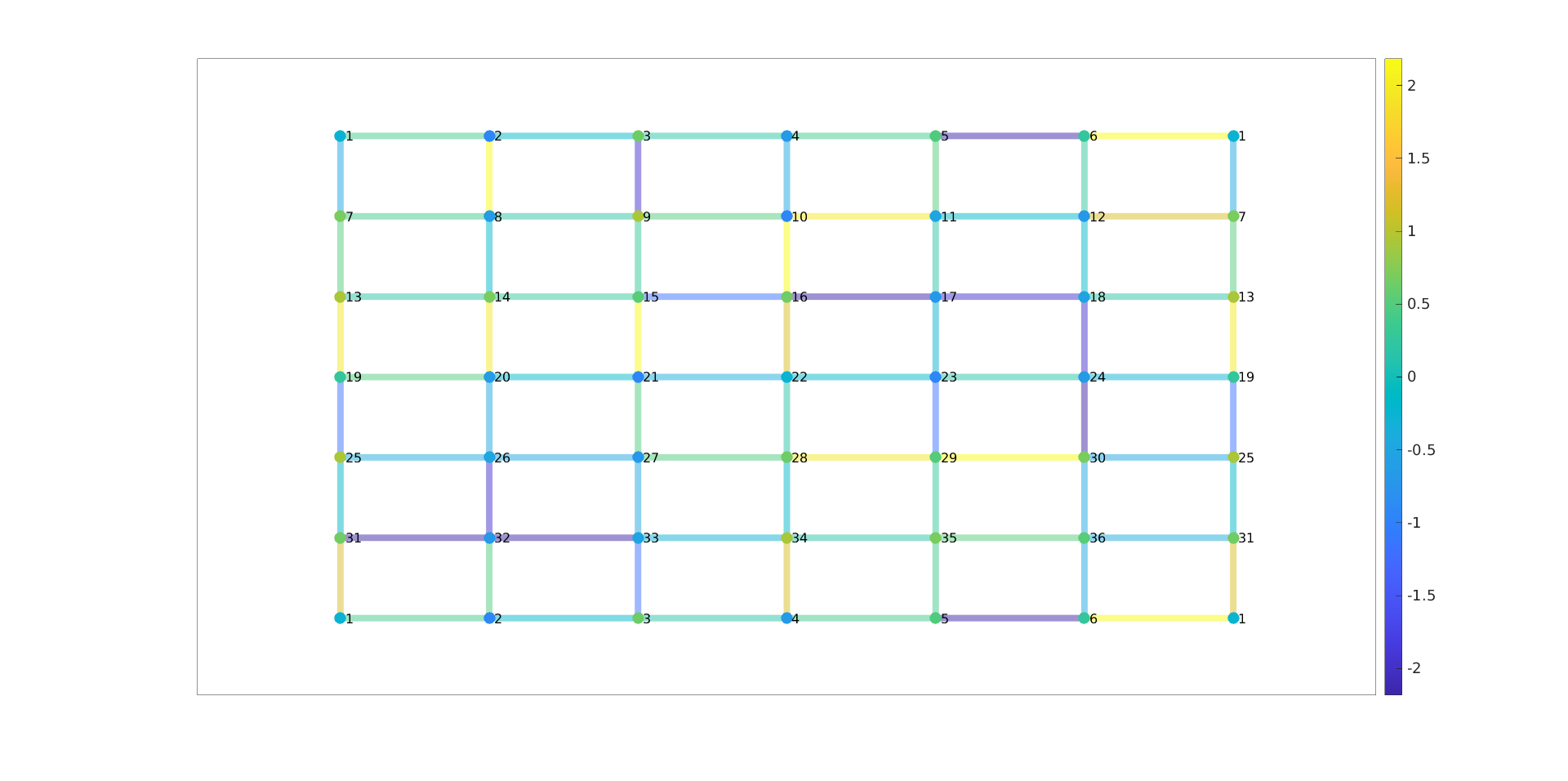}
\caption{\textbf{Turing patterns on the 2D lattice with periodic boundary conditions} Snapshot, at $60$ time units, of the patterns for species $u$ (nodes) and $w$ (links) visualized on the network. Such dynamics can be better appreciated in the video (see the Supplementary Material of the online version of this paper). The periodicity of the lattice can be visualized thanks to the node labeling. The parameters are $\sigma_1=1,~\sigma_2=-1.4,~\sigma_3=-2,~\eta_1=-3,~\xi_1=-0.5,~\xi_2=5,
~\zeta_1=0.1,~\zeta_2=6,~\zeta_3=6,~\zeta_4=1,~c_1=-1,~c_2=1,~\alpha_u=1,~\alpha_v=1,~\beta_u=1,~\beta_v=0.1$; the initial perturbation is the order of $10^{-2}$.}
\label{fig:Turing_patterns_2D_lattice}
\end{figure}

We now compare the dynamical patterns observed in presence of diffusion ($c_2\neq0$) and in the case when it is absent ($c_2=0$).
We chose  parameters values such that, in principle, dynamical (oscillatory) Turing patterns are obtained, i.e., parameters in blue region. 
Let us first discuss the phenomenology of the dynamical Turing patterns observed in presence of diffusion.
The dynamical nature of these topological patterns is revealed by the dispersion relation  displayed in Fig. \ref{fig:rel_disp_time_series_Tur}a), where we can observe that the mode associated to $\mu_k=0$ is stable (indeed $\lambda(0)<0$), while there exist some modes that are unstable ($\lambda(\mu_k)>0$ for some $k$) driving the formation of Turing patterns. In panel b) of the same figure, we can observe that the imaginary part of the dispersion relation $\varrho$ is non-zero for corresponding unstable modes. For this reason, the obtained Turing patterns on nodes and links are dynamical, i.e., oscillatory, as it is apparent from  Fig. \ref{fig:rel_disp_time_series_Tur}c), where we display the time series for the two species on the nodes, i.e., $u$ and $v$, and for the species on the links, i.e., $w$. Let us point out that the form of Turing patterns is perturbation dependent and different perturbation yield different (oscillatory) patterns. Lastly, in Fig. \ref{fig:Turing_patterns_2D_lattice}, we display a snapshot of the patterns for species $u$ on the nodes and $w$ on the links visualized on the network\footnote{Such dynamics can be better appreciated in the video 
available as Supplementary Material of the online version.}.

\begin{figure}[ht!]
\centering
\includegraphics[scale=1]{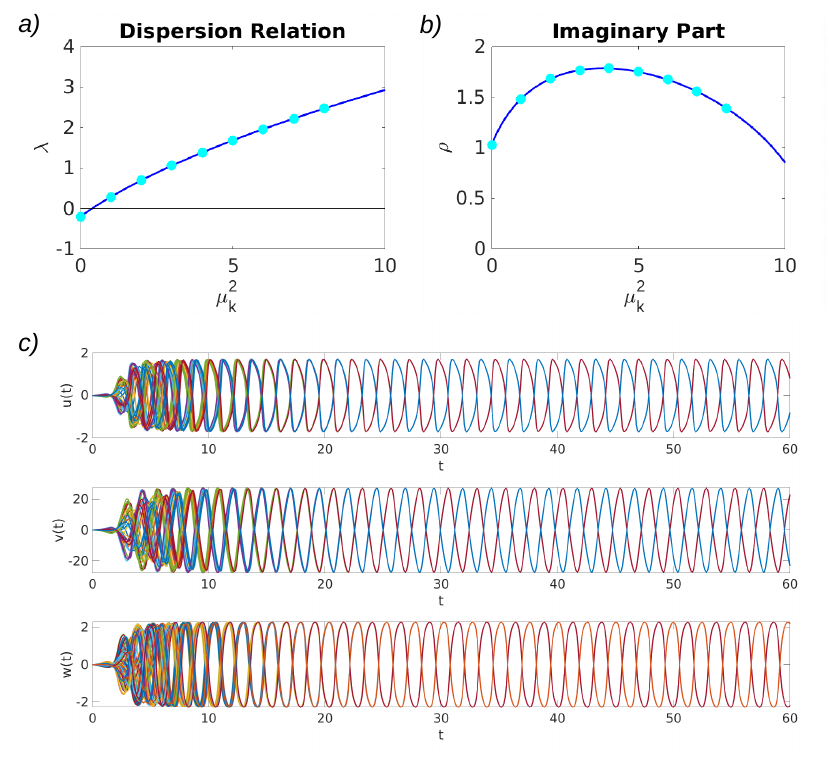}
\caption{\textbf{Dispersion relation and Dirac-induced patterns}
Panel a) shows the real part of the dispersion relation $\lambda$, while panel b) its respective imaginary part $\varrho$. The dispersion relation is computed as for the case with diffusion (Fig. \ref{fig:rel_disp_time_series_Tur}, panels a and b). We can observe that the real part of the dispersion relation is divergent, however this does not pose a problem given that our support is discrete and finite, and so is its spectrum. The chosen setting is such that Dirac-induced patterns emerge ($\lambda>0$) and they are oscillatory (the corresponding $\varrho\neq 0$), as shown by the dynamics of the $3$ interacting species, in panel c). The network is a $2$D lattice with periodic boundary conditions of $36$ nodes and the parameters are $\sigma_1=1,~\sigma_2=-1.4,~\sigma_3=-10,~\eta_1=-3,~\xi_1=-0.5,~\xi_2=5,
~\zeta_1=0.1,~\zeta_2=6,~\zeta_3=6,~\zeta_4=1,~c_1=-1,~c_2=0,~\alpha_u=1,~\alpha_v=1,~\beta_u=1,~\beta_v=0.1$; the initial perturbation is the order of $10^{-2}$.}
\label{fig:rel_disp_dir_time_series}
\end{figure}

\begin{figure}[ht!]
\centering
\includegraphics[scale=0.3]{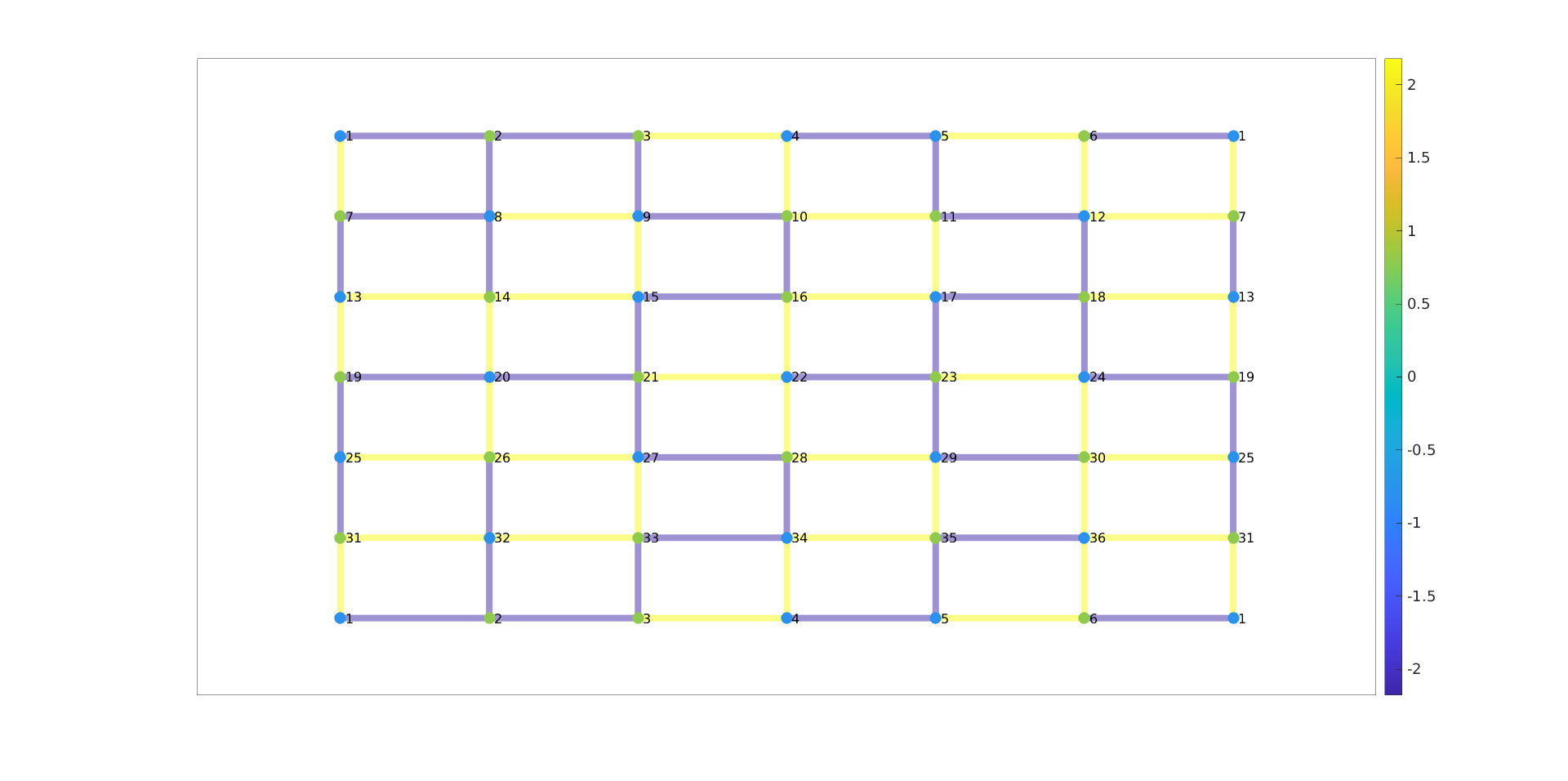}
\caption{\textbf{Dirac-induced patterns on the 2D lattice with periodic boundary conditions} Snapshot, at $60$ time units, of the patterns for species $u$ (nodes) and $w$ (links) visualized on the network. Such dynamics can be better appreciated in the video (see the Supplementary Material of the online version of this paper). The parameters are $\sigma_1=1,~\sigma_2=-1.4,~\sigma_3=-10,~\eta_1=-3,~\xi_1=-0.5,~\xi_2=5,
~\zeta_1=0.1,~\zeta_2=6,~\zeta_3=6,~\zeta_4=1,~c_1=-1,~c_2=0,~\alpha_u=1,~\alpha_v=1,~\beta_u=1,~\beta_v=0.1$; the initial perturbation is the order of $10^{-2}$.}
\label{fig:Dirac_patterns_2D_lattice}
\end{figure}

We now compare the phenomenology of the Turing patterns obtained in presence of diffusion, i.e., $c_2>0$, with the ones arising once solely using the Dirac operator, i.e., $c_2=0$.
Also in this case, we consider a set of parameters for which the stability of the homogeneous equilibrium is lost once perturbed, as revealed by the real part of the dispersion relation in Fig. \ref{fig:rel_disp_dir_time_series}a). We can also remark that, for this choice of parameters, the latter is divergent for large $\mu_k$. When working on continuous support, where all instability modes are present, this is a problem because the maximum critical mode has an infinite wave number, hence the solution is not a physical one (long wave instability). However, on discrete support the finite size of the latter induces an upper bound on the largest possible Laplace eigenvalue. In particular it is well known that on networks the maximum eigenvalue of the $0$-Laplacian (and the isomorphic $1$-Laplacian) is bounded for networks with bounded degrees. Finally, even  for networks in with unbounded degrees it is possible to consider a bounded dispersion relation by adopting the normalized Dirac~\cite{baccini2022weighted}  and  Laplacian operators. A further discussion of the dispersion relation for large $\mu_k$ can be found in Appendices \ref{app:A} and \ref{app:B}.
By considering the obtained dispersion relation we note that also in this case, as in the previous one, its imaginary part $\varrho$ is non-zero for the unstable modes (see Figure \ref{fig:rel_disp_dir_time_series}, panel b)); the resulting dynamical patterns can be visualized in panel c). Lastly, in Fig. \ref{fig:Dirac_patterns_2D_lattice}, we display a snapshot of the patterns for species $u$ on the nodes and $w$ of the links visualized on the $2$D lattice\footnote{The dynamics can be better appreciated in the video 
available as Supplementary Material of the online version.}. Remarkably, Dirac-induced patterns present several distinct dynamical properties that are not present in Turing pattern observed for $c_2\neq 0$.

\begin{figure}[ht!]
\centering
\includegraphics[scale=1]{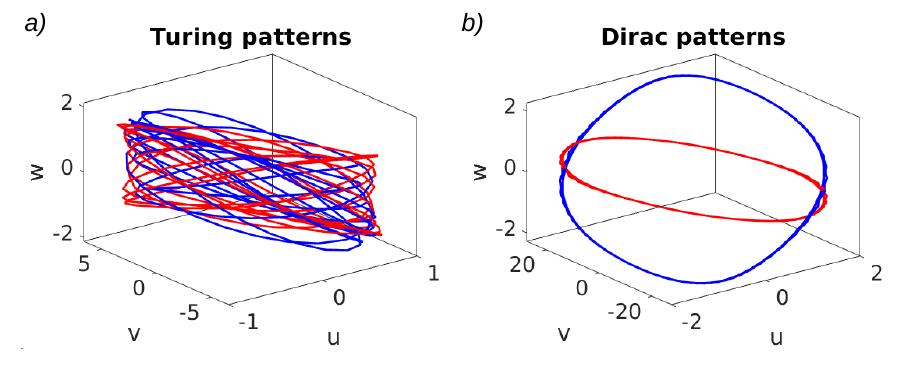}
\caption{\textbf{Projected attractors} a) Turing patterns, b) Dirac-induced patterns. The network is a $2$D lattice with periodic boundary conditions of $36$ nodes. The parameters are $\sigma_1=1,~\sigma_2=-1.4,~\eta_1=-3,~\xi_1=-0.5,~\xi_2=5,
~\zeta_1=0.1,~\zeta_2=6,~\zeta_3=6,~\zeta_4=1,~c_1=-1,~\alpha_u=1,~\alpha_v=1,~\beta_u=1,~\beta_v=0.1$; for panel a) we have $\sigma_3=-2,~c_2=1$, while for panel b) $\sigma_3=-10,~c_2=0$; the initial perturbation is the order of $10^{-2}$.}
\label{fig:attractor_3D}
\end{figure}

In particular we observe that nodes and links are phase-synchronized and divided in two clusters (see Fig. \ref{fig:rel_disp_dir_time_series}c)), which recalls a phenomenon called \textit{cluster synchronization}, often occurring when the network possesses symmetries \cite{pecora2014cluster}, as it is for our case. Moreover, such patterns do not change when varying the initial conditions, at contrast with Turing patterns. 
To better visualize the difference in the behavior of the two dynamical patterns, Laplace-driven and Dirac-induced, respectively, let us project the dynamics in the $3$D phase space $(u,v,w)$. The results are shown in Fig. \ref{fig:attractor_3D}, with panel a) depicting the ``attractor'' for Laplace-driven patterns, while panel b) for Dirac-induced patterns. More precisely, we select one link, $\ell$, we consider one of its end-node, $i$, and we build the orbit $(u_i(t),v_i(t),w_\ell(t))$. Let us observe that, while a representation in the $(u,v)$ phase space is unique, the shown representation is not, because there are more links thank nodes and there is an ambiguity in the selection of the end-node for each link. For what concerns Dirac-induced patterns, however, even by reshuffling the $w$ in the triplets, the attractor does not change. This is different for Turing patterns and, indeed, the $3$D attractor as the sole scope of showing the difference between the two cases, rather than an accurate representation of the former. An important fact is that the Dirac attractor is robust with respect to the initial condition and all the trajectories converge to it, no matter the size of the initial perturbation. Such behavior is remarkably different from the one arising with Turing patterns, which are highly sensible to even small variations on the initial conditions.

\subsection{Phenomenology on a random graph}
\label{sec:phen}

\begin{figure}[ht!]
\centering
\includegraphics[scale=0.4]{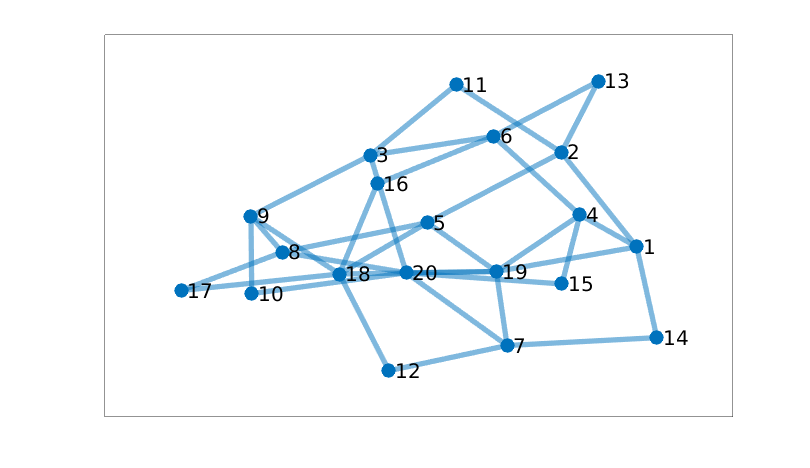}
\caption{\textbf{Random Eulerian Network} The considered random Eulerian network of $N_0=20$ nodes with degree distribution $P(2)=8/20$, $P(4)=9/20$, $P(6)=3/20$ and $P(k)=0$ for all $k\neq 2,4,6$.}
\label{fig:ren}
\end{figure}

We further investigate the nature and the phenomenology of the patterns by analyzing the case of a random Eulerian network, shown in Fig. \ref{fig:ren}. It is important to deal with a random structure, so we can better understand what causes the regularity of the Dirac-induced pattern on the lattice. 
Specifically we consider a random network with $N_0=20$ nodes and degree distribution  $P(2)=8/20$, $P(4)=9/20$, $P(6)=3/20$ and $P(k)=0$ for all $k\neq 2,4,6$. Specifically this network has all the nodes with even degree, hence it is Eulerian. This implies that if we label the nodes subsequently along any of its Eulerian paths and we orient the links according to the node label, then it is immediate to show that the network admits a constant eigenvector ${\bf 1}$ in the kernel of the Dirac operator.

\begin{figure}[ht!]
\centering
\includegraphics[scale=1]{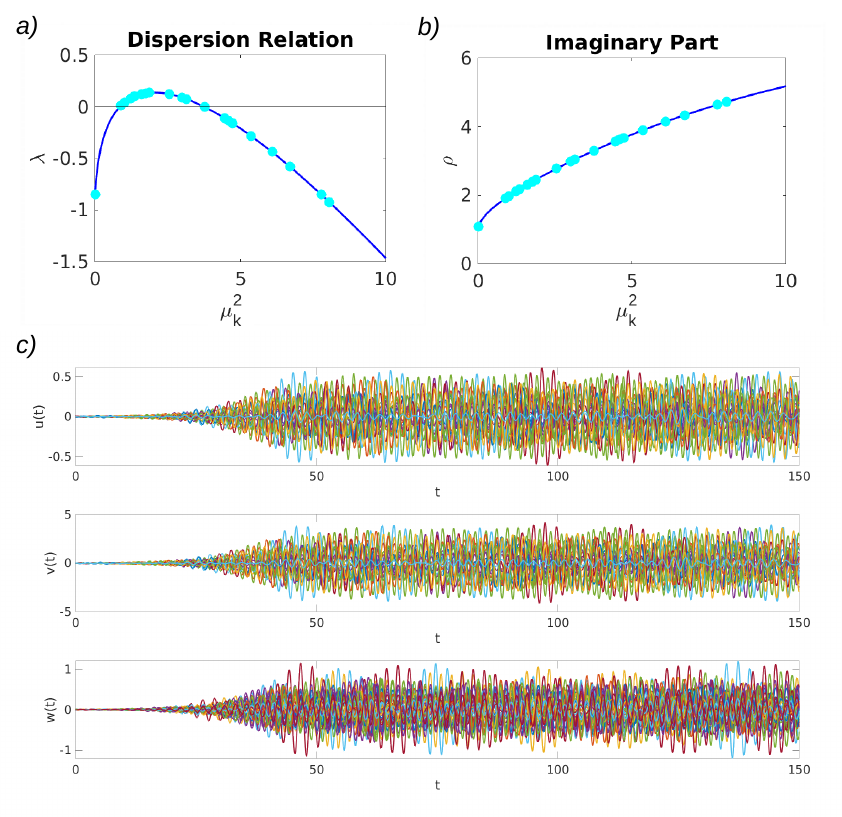}
\caption{\textbf{Dispersion relation and Turing patterns} Panel a) shows the dispersion relation $\lambda$, while panel b) its associated imaginary part $\varrho$. As for the case of $2$D lattice, the chosen setting is such that Dirac-induced patterns emerge ($\lambda>0$) and they are oscillatory (the corresponding $\varrho\neq 0$), as shown by the dynamics of the $3$ interacting species, in panel c).
 The network is that of Fig. \ref{fig:ren} and the parameters are $\sigma_1=1,~\sigma_2=-1.4,~\sigma_3=-2,~\eta_1=-3,~\xi_1=-0.5,~\xi_2=5,
~\zeta_1=0.1,~\zeta_2=6,~\zeta_3=6,~\zeta_4=1,~c_1=-1,~c_2=1,~\alpha_u=1,~\alpha_v=1,~\beta_u=1,~\beta_v=0.1$; the initial perturbation is the order of $10^{-2}$.}
\label{fig:rel_disp_ren_time_series}
\end{figure}

\begin{figure}[ht!]
\centering
\includegraphics[scale=1]{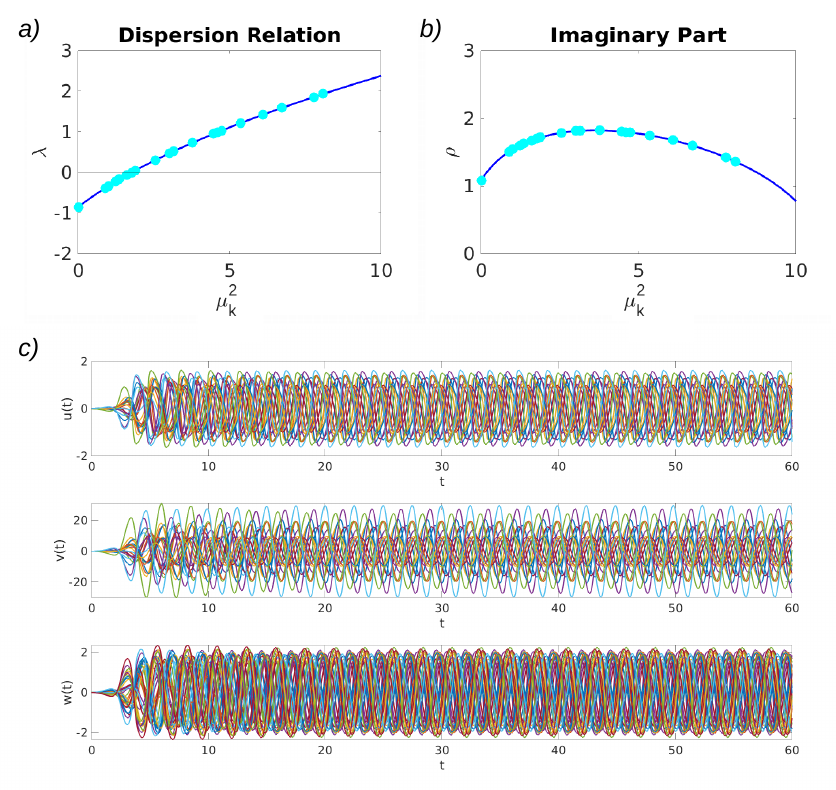}
\caption{\textbf{Dispersion relation and Turing patterns} Panel a) shows the real part of the dispersion relation $\lambda$, while panel b) its respective imaginary part $\varrho$. As for the case of $2$D lattice, the chosen setting is such that Dirac-induced patterns emerge ($\lambda>0$) and they are oscillatory (the corresponding $\varrho\neq 0$), as shown by the dynamics of the $3$ interacting species, in panel c).
 The network is that of Fig. \ref{fig:ren} and the parameters are $\sigma_1=1,~\sigma_2=-1.4,~\sigma_3=-10,~\eta_1=-3,~\xi_1=-0.5,~\xi_2=5,
~\zeta_1=0.1,~\zeta_2=6,~\zeta_3=6,~\zeta_4=1,~c_1=-1,~c_2=0,~\alpha_u=1,~\alpha_v=1,~\beta_u=1,~\beta_v=0.1$; the initial perturbation is the order of $10^{-2}$.}
\label{fig:rel_disp_dir_ren_time_series}
\end{figure}

In order to investigate the dynamical properties of the model  \eqref{eq:rd_2nod1link_model} on this random topology, we distinguish between Turing patterns, i.e., once the Laplace matrix is present, and Dirac-induced patterns, once the coupling is solely realized by using the Dirac operator. In both cases we consider parameter values for which the homogeneous equilibrium becomes unstable and (in principle) oscillatory.
 In Fig.  \ref{fig:rel_disp_dir_ren_time_series}(a-b), we plot the dispersion relation corresponding to this choice of parameter values and, in panel c), we display the resulting dynamical Turing patterns. These are qualitatively similar to those obtained on the lattice network, with the only difference that the transient is longer. The latter is due to the fact that, for this choice of parameters, the critical mode is closer to zero (compare Fig. \ref{fig:rel_disp_ren_time_series}a) with \ref{fig:rel_disp_time_series_Tur}a))\footnote{By comparing the Figures, one can observe that the spectral gap of the random Eulerian network is larger, hence motivating a slightly different choice of parameters. Let us recall that, on networks, the conditions for instability are necessary but not sufficient.}.

\begin{figure}[ht!]
\centering
\includegraphics[scale=1]{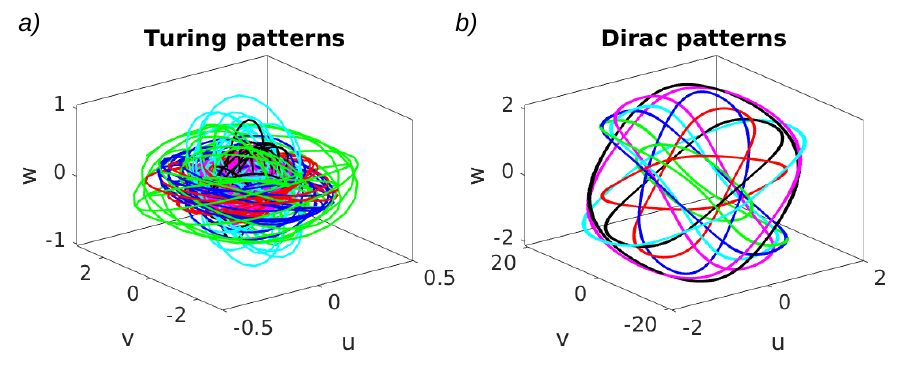}
\caption{\textbf{Projected attractors} a) Turing patterns, b) Dirac-induced patterns. The network is the Random Eulerian Network of Fig. \ref{fig:ren}. a) The parameters are $\sigma_1=0.3,~\sigma_2=-2,~\sigma_3=-2,~\eta_1=-3,~\xi_1=-0.5,~\xi_2=5,
~\zeta_1=0.1,~\zeta_2=6,~\zeta_3=6,~\zeta_4=1,~c_1=-1,~c_2=1,~\alpha_u=1,~\alpha_v=1,~\beta_u=1,~\beta_v=0.1$. b) The parameters are $\sigma_1=0.3,~\sigma_2=-2,~\sigma_3=-10,~\eta_1=-3,~\xi_1=-0.5,~\xi_2=5,
~c_1=-1,~\alpha_u=1.1,~\alpha_v=7,~\beta_u=7,~\beta_v=1.1$; the initial perturbation is the order of $10^{-2}$.}
\label{fig:attractor_3D_ren}
\end{figure}

Let us now repeat the numerical analysis for the case without diffusion, i.e., $c_2=0$. Again, the chosen setting yields oscillatory patterns, i.e., parameters in the blue region of Fig. \ref{fig:Tur_Dir_regions}b), as can be appreciated by looking at the dispersion relation, depicted in Fig. \ref{fig:rel_disp_dir_ren_time_series}(a-b); the resulting Dirac-induced patterns can be visualized in panel c). We can observe that they are not clustered as for the case on lattice, but we can somehow spot some kind of regularity. We corroborate this qualitative observation by plotting the trajectories in the phase space, shown in Fig. \ref{fig:attractor_3D_ren}, for Turing patterns (panel a) and Dirac-induced patterns (panel b). There, we can see that Turing patterns are qualitatively equivalent to those obtained on the lattice network; moreover, as expected, they are not robust with respect to the initial perturbation. On the contrary, Dirac-induced patterns are, again, robust to the initial conditions and their regularity can be visualized through the closed curves in the phase space. 

These observations allow us to conclude that Dirac-induced patterns are robust with respect to the initial conditions and, might be sensible to the symmetries of the underlying network topology.

\section{Conclusions}\label{sec:concl}

The topological approach to network dynamics implies that the dynamical state of a network is encoded in the direct sum of topological signals on nodes ($0$-cochains) and topological signals on links ($1$-cochains). This approach significantly changes our description of dynamical processes on networks.
Topological signals on nodes and links can be coupled by the Dirac operator that represents the ``square-root" of the Laplacian.
Here we have shown that dynamical Turing and Dirac-induced patterns can set up on nodes and links of Eulerian networks as a result of the instability of the homogeneous steady state solution. This approach extends previous results on node-based Turing patterns defined on networks, simplicial complexes and time-varying networks.
Let us observe that the setting where a single dynamical variable is associated to nodes and a single dynamical variable is associated to links cannot account for dynamical Turing and Dirac patterns \cite{turing_topological}. Here we have shown that taking three topological signals (two on the nodes and one on the links) can instead account for dynamical patterns on nodes and links. In order to describe their dynamics we have defined the $3$-way Hodge-Dirac operator that is projecting the node signals into the links and, vice versa, is projecting the link signals into the nodes. This operator is then combined to an appropriate gamma matrix that has the ability to compress the projected node signals into a single link signal and expand the projected link signal into two distinct node signals.
In the absence of coupling, the topological signals of each dimension obey the same dynamical equation, while the coupling can be induced by either the higher-order Laplacian matrix or the $3$-way Dirac operator, or a combination of both operators. We called Dirac-induced patterns the ones induced exclusively by the Dirac operator, while we call all the other patterns Turing patterns.
The formulated theory of Turing and Dirac induced patterns leads to the instability of the uniform steady state solution and the onset of a dynamical behavior that displays remarkable differences observed for a significant region of the parameter space.
The difference between the observed dynamical behavior of Turing and Dirac induced patterns is here highlighted by considering two very distinct topologies: a $2$-dimensional torus tessellated by a square lattice and a random (Eulerian) graph.
In both cases, it emerges that the Dirac induced patterns have distinct dynamical properties displaying a clustering of the dynamics characterized by an increased robustness to noise with respect to Turing patterns.

We believe that this work can open new scenarios in the treatment of topological Turing instabilities affect jointly nodes and links signals with possible applications to neuroscience.

\section*{Author contribution statement}
\noindent Riccardo Muolo: Conceptualization, Formal analysis, Methodology, Software, Investigation,  Writing - Original draft preparation\\
\noindent Timoteo Carletti: Formal analysis, Investigation, Writing - Original draft preparation\\
\noindent Ginestra Bianconi: Conceptualization, Formal analysis, Methodology, Investigation, Writing - Original draft preparation

\section*{Declaration of competing interest}
The authors declare that they have no known competing financial interests or personal relationships that could have appeared to influence the work reported in this paper.

\section*{Data availability}
Data will be made available upon reasonable request to the corresponding author.

\section*{Funding}
R.M. acknowledges JSPS KAKENHI JP22K11919, JP22H00516, and JST CREST JP-MJCR1913 for financial support. During part of this work, R.M. was supported by a FRIA-FNRS Fellowship, funded by the Walloon Region, Grant FC 33443.

\section*{Acknowledgements}
The authors are grateful to Benedetta Franceschiello, Lorenzo Giambagli and Lucille Calmon for interesting discussions.

\appendix

\section{Conditions for the emergence of Turing patterns}
\setcounter{equation}{0}
\renewcommand{\theequation}{A\arabic{equation}}
\setcounter{figure}{0}
\renewcommand{\thefigure}{A\arabic{figure}}\label{app:A} 

We can determine the conditions to obtain Turing patterns by studying the stability of the following polynomial

\begin{equation}\label{eq:apppoly}
    a \lambda_k^3 + b\lambda_k^2 + c\lambda_k +d = 0,
\end{equation}

whose coefficients are

\begin{displaymath}
\begin{cases}
    a=1, \\
    b=c_2\mu_k^2(D_{uu}+D_{vv}+D_{ww})-(f_{u}+g_{v}+h_{w}),\\
    c= \pi_1(\mu_k^2),\\
    d= \pi_2(\mu_k^2),
    \end{cases}
\end{displaymath}
where $\pi_1$ and $\pi_2$ are polynomials in $\mu_k^2$. \\

The first polynomial has degree $2$ and has the following coefficients $\pi_1(\mu_k^2)=\mathcal{A} \mu_k^4+ \mathcal{B} \mu_k^2 +\mathcal{C}$, explicitly
\begin{displaymath}
\begin{cases}
    \mathcal{A}=(D_{vv}D_{ww}+D_{uu}D_{ww}+D_{uu}D_{vv}-D_{uv}D_{vu})c_2^2, \\
    \mathcal{B}=[D_{uv}g_u+D_{vu}f_v-D_{uu}(h_w+g_v)+\\
    ~~~~~~~~-D_{vv}(h_w+f_u)-D_{ww}(f_u+g_v)]c_2-(f_wh_u+g_wh_v),\\
    \mathcal{C}= f_ug_v-f_vg_u+h_w(f_u+g_v),
    \end{cases}
\end{displaymath} while the second one has degree $3$ and coefficients are $\pi_2(\mu_k^2)=\mathcal{E} \mu_k^6+ \mathcal{F} \mu_k^4 +\mathcal{G}\mu_k^2+\mathcal{H}$, explicitly

\begin{displaymath}
\begin{cases}
    \mathcal{E}= D_{ww}(D_{uu} D_{vv}-D_{uv} D_{vu}){c_{2}}^3 ,\\
    \mathcal{F}=[f_w(h_vD_{vu}-h_uD_{vv})+g_w(h_uD_{uv}-h_vD_{uu})]c_2+\\
    ~~~~~~~~~~~~~~~~~~~~~~~~~~+[D_{ww}(f_vD_{vu}-f_uD_{vv}+g_uD_{uv}-g_vD_{uu})+h_w(D_{uu}D_{vv}+D_{uv}D_{vu})]c_2^2,\\
    \mathcal{G}=g_w(f_uh_v-f_vh_u)+f_v(g_vh_u-g_uh_v)+[D_{ww}(f_ug_v-f_vg_u)+h_w(f_uD_{vv}+g_vD_{uu}-f_vD_{vu}-g_uD_{uv}]c_2.\\
    \mathcal{H}=h_w(f_{v} g_{u}-f_{u} g_{v})
    \end{cases}
\end{displaymath}

Let us observe that \begin{displaymath} 
D_{uu} D_{vv}-D_{uv} D_{vu}=\alpha_u\beta_u\alpha_v\beta_v-\alpha_v\beta_v\alpha_v\beta_u=0,
\end{displaymath}
hence $\mathcal{E}=0$ and $\pi_2(\mu_k^2)$ is \textit{de facto} a polynomial of degree $2$, i.e., $\pi_2(\mu_k^2)=\mathcal{F} \mu_k^4 +\mathcal{G}\mu_k^2+\mathcal{H}$.\\

The Routh-Hurwitz criterion \cite{Routh1877,Hurwitz1895} gives us the following necessary and sufficient conditions for stability \begin{equation}\label{eq:appcond2}
    \begin{aligned}
        a>&~0 , \\
        b>&~0 , \\
bc-ad>&~0 , \\
d>&~0 .
\end{aligned} 
\end{equation}

We are interested in the conditions to obtain Turing patterns, i.e., at least one of the roots of the polynomial has a positive real part. This means that it is enough that one of the \eqref{eq:appcond2} is violated. \\

The first and second conditions cannot be violated: $a>0$ is trivial, while $b>0$ needs to be satisfied due to the stability of the homogeneous equilibrium, which gives us that $f_u + g_v<0$ and $h_w<0$. Let us hence study the fourth condition $d>0$. Namely, we need check if the polynomial $\pi_2(\mu_k^2)=\mathcal{F} \mu_k^4 +\mathcal{G}\mu_k^2+\mathcal{H}$ can take negative values. We can observe that $\mathcal{H}=h_w(f_{v} g_{u}-f_{u} g_{v})>0$, again for the conditions on the homogeneous equilibrium. Then we need that either $\mathcal{F}<0$ (so that we are sure that $\pi_2$ takes some negative values, independently of $\mathcal{G}$) or $\mathcal{G}<0$ ($\pi_2$ may be negative, independently of $\mathcal{F}$, but not for every set of parameters). The same principles apply to imposing the negativity of the third of the Routh-Hurwitz conditions \eqref{eq:appcond2}, i.e., $bc-ad<0$, which can be obtained in an analogous way. In the main text we put ourselves in a setting in which Eqs. \eqref{eq:appcond2} are violated and hence the system exhibits Turing patterns, which are, as expected, dynamics.

Let us conclude this section by studying the asymptotic behavior of the roots of the polynomial~\eqref{eq:apppoly} for $\mu_k \gg 1$. The first step is to rewrite the latter as follows
\begin{equation*}
p(\lambda) = \mu_k^4 A(\lambda)+\mu_k^2 B(\lambda)+C(\lambda)\, ,
\end{equation*}
where $A$, $B$ and $C$ are polynomials in $\lambda$ of degree $1$, $2$ and $3$. By dividing the previous expression by $\mu_k^4$ we obtain a second polynomial in $\lambda$ with the same roots as the previous one. Let us first consider $\mu_k \gg 1$ and write the root of the polynomial as $\lambda(\mu_k)=\lambda_0 + \lambda_1/\mu_k^2 +\mathcal{O}(\mu_k^{-4})$. Then, it follows that $\lambda_0$ satisfies $A(\lambda_0)=0$ and the second term $\lambda_1$ can be obtained by solving $A'(\lambda_0)\lambda_1+B(\lambda_0)=0$. Because $A$ is a linear function, its root can be straightforwardly determined
\begin{eqnarray}
\lambda_0 = -\frac{1}{c_2 (\alpha_u \beta_u + \alpha_v \beta_v)^2} &\Big[ &\alpha_v\beta_v f_w h_u - \alpha_v \beta_u f_w h_v + \alpha_v^2 \beta_v^2 c_2 f_u - \alpha_v^2 \beta_u \beta_v c_2 f_v - \alpha_u \beta_v g_w h_u \nonumber \\
&+&   \alpha_u \beta_u g_w h_v - \alpha_u \alpha_v \beta_v^2 c_2 g_u +
\alpha_u \alpha_v \beta_u \beta_v c_2 g_v + \alpha_u \alpha_v \beta_u \beta_v c_2 f_u \nonumber \\
&-&  \alpha_u \alpha_v \beta_u^2 c_2 f_v-\alpha_u^2 \beta_u \beta_v c_2 g_u + \alpha_u^2 \beta_u^2 c_2 g_v\Big]. \nonumber 
\end{eqnarray}
This implies that for sufficiently large $\mu_k$ the roots of~\eqref{eq:apppoly} will be close to $\lambda_0$ within a term of order $1/\mu_k^2$.

\section{Conditions for the emergence of Dirac-induced patterns}\label{app:B}
\setcounter{equation}{0}
\renewcommand{\theequation}{B\arabic{equation}}
\setcounter{figure}{0}
\renewcommand{\thefigure}{B\arabic{figure}}

To compute the conditions yielding the emergence of Dirac-induced patterns, one can proceed in the same way as before. Moreover, let us remark that the conditions for the stability of the homogeneous equilibrium are the same of the previous case, of the uncoupled dynamics has not changed. 

The fact that $c_2=0$ simplifies the coefficients of the polynomial whose stability needs to be determined. In fact, now we are dealing with the following:

\begin{equation}\label{eq:apppoly2}
    a \lambda_k^3 + b\lambda_k^2 + c\lambda_k +d = 0,
\end{equation}

where

\begin{displaymath}
\begin{cases}
    a=1 \\
    b=-(f_{u}+g_{v}+h_{w}),\\
    c= \varrho_1(\mu_k^2),\\
    d= \varrho_2(\mu_k^2),
    \end{cases}
\end{displaymath}
and $\varrho_1$ and $\varrho_2$ are now linear curves in $\mu_k^2$. \\

The first one has the following form $\varrho_1(\mu_k^2)=\mathcal{X} \mu_k^2 +\mathcal{Y}$, explicitly
\begin{displaymath}
\begin{cases}
    \mathcal{X}= -(f_wh_u+g_wh_v),\\
    \mathcal{Y}= f_ug_v-f_vg_u+h_w(f_u+g_v),
    \end{cases}
\end{displaymath} while the second one is  $\varrho_2(\mu_k^2)=\mathcal{Z}\mu_k^2+\mathcal{K}$, explicitly:

\begin{displaymath}
\begin{cases}
    \mathcal{Z}=g_w(f_uh_v-f_vh_u)+f_v(g_vh_u-g_uh_v),\\
    \mathcal{K}=h_w(f_{v} g_{u}-f_{u} g_{v}).
    \end{cases}
\end{displaymath}

The Routh-Hurwitz criterion \cite{Routh1877,Hurwitz1895} gives us the following necessary and sufficient conditions for stability \begin{equation}\label{eq:appcond}
    \begin{aligned}
        a>&~0,  \\
        b>&~0,  \\
bc-ad>&~0  ,\\
d>&~0 .
\end{aligned} 
\end{equation}

As before, $a$ and $b$ are always positive and such conditions can never be violated. Let us hence study the condition $d<0$, i.e., we need to impose that $\varrho_1<0$. Since the coefficient $\mathcal{K}$ is always positive, the condition reduces to imposing $\mathcal{Z}$ negative 

\begin{equation}\label{eq:cond1_app}
    g_w(f_uh_v-f_vh_u)+f_v(g_vh_u-g_uh_v)<0,
\end{equation}

which is our first condition for Dirac-induced patterns. Let us then proceed in studying the Routh-Hurwitz condition $bc-ad<0$, which translates into a new linear curve $\varrho_3(\mu_k^2)=\mathcal{J}\mu_k^2+\mathcal{L}$, where now \begin{displaymath}
    \begin{cases}
        \mathcal{J}=b\mathcal{X}-\mathcal{Z}=(f_{u}+g_{v}+h_{w})(f_wh_u+g_wh_v)-g_w(f_uh_v-f_vh_u)-f_v(g_vh_u-g_uh_v), \\
        \mathcal{L}=b\mathcal{Y}- \mathcal{K}= -(f_{u}+g_{v}+h_{w})[f_ug_v-f_vg_u+h_w(f_u+g_v)]-h_w(f_{v} g_{u}-f_{u} g_{v}).
    \end{cases}
\end{displaymath}

Hence, patterns are obtained if $\mathcal{J}<0$ or if $\mathcal{L}<0$, namely if at least one the two condition
\begin{equation}\label{eq:cond2_app}
\begin{cases}
   (f_{u}+g_{v}+h_{w})(f_wh_u+g_wh_v)<g_w(f_uh_v-f_vh_u)+f_v(g_vh_u-g_uh_v),\\
    (f_{u}+g_{v}+h_{w})[f_ug_v-f_vg_u+h_w(f_u+g_v)]>h_w(f_{v} g_{u}-f_{u} g_{v}),
    \end{cases}
\end{equation}
is satisfied. Let us observe that the second condition is never satisfied because of the conditions for the homogeneous equilibrium. In fact, the expression $ (f_{u}+g_{v}+h_{w})[f_ug_v-f_vg_u+h_w(f_u+g_v)]$ is always negative, while $h_w(f_{v} g_{u}-f_{u} g_{v})$ is always positive. \\ 

Summing up, the conditions for Dirac-induced patterns are the following

\begin{equation}\label{eq:cond_final_app}
\begin{cases}
    g_w(f_uh_v-f_vh_u)+f_v(g_vh_u-g_uh_v)<0,\\
   (f_{u}+g_{v}+h_{w})(f_wh_u+g_wh_v)<g_w(f_uh_v-f_vh_u)+f_v(g_vh_u-g_uh_v).
    \end{cases}
\end{equation}

Also in this case, we can study the asymptotic behavior of the roots of the polynomial~\eqref{eq:apppoly2} for $\mu_k \gg 1$. Let us thus rewrite the latter polynomial as follows
\begin{equation*}
p(\lambda) = \mu_k^2 \tilde{B(}\lambda)+\tilde{C}(\lambda)\, ,
\end{equation*}
where $\tilde{B}$ and $\tilde{C}$ are polynomials in $\lambda$ of degree $1$ and $3$. By dividing the previous expression by $\mu_k^2$, we obtain again a polynomial in $\lambda$ sharing the same roots as the previous one. Let us assume $\mu_k \gg 1$ and write the root of the polynomial as $\lambda(\mu_k)=\lambda_0 + \lambda_1/\mu_k^2 +\mathcal{O}(\mu_k^{-4})$. Then, $\lambda_0$ satisfies $\tilde{B}(\lambda_0)=0$ and the second term $\lambda_1$ can be obtained by solving $\tilde{B}'(\lambda_0)\lambda_1+\tilde{C}(\lambda_0)=0$. Being $\tilde{B}$ a first degree polynomial, its root can be straightforwardly determined
\begin{equation*}
\lambda_0 = -\frac{f_w g_v h_u - f_w g_u h_v - f_v g_w h_u + f_u g_w h_v}{g_w h_v + f_w h_u}
\, .
\end{equation*}
This implies that for sufficiently large $\mu_k$ the roots of~\eqref{eq:apppoly2} will be close to $\lambda_0$ within a correction of order $1/\mu_k^2$.

\section{Stationary patterns}\label{app:C}
\setcounter{equation}{0}
\renewcommand{\theequation}{C\arabic{equation}}
\setcounter{figure}{0}
\renewcommand{\thefigure}{C\arabic{figure}}

\begin{figure}[ht!]
\centering
\includegraphics[scale=0.85]{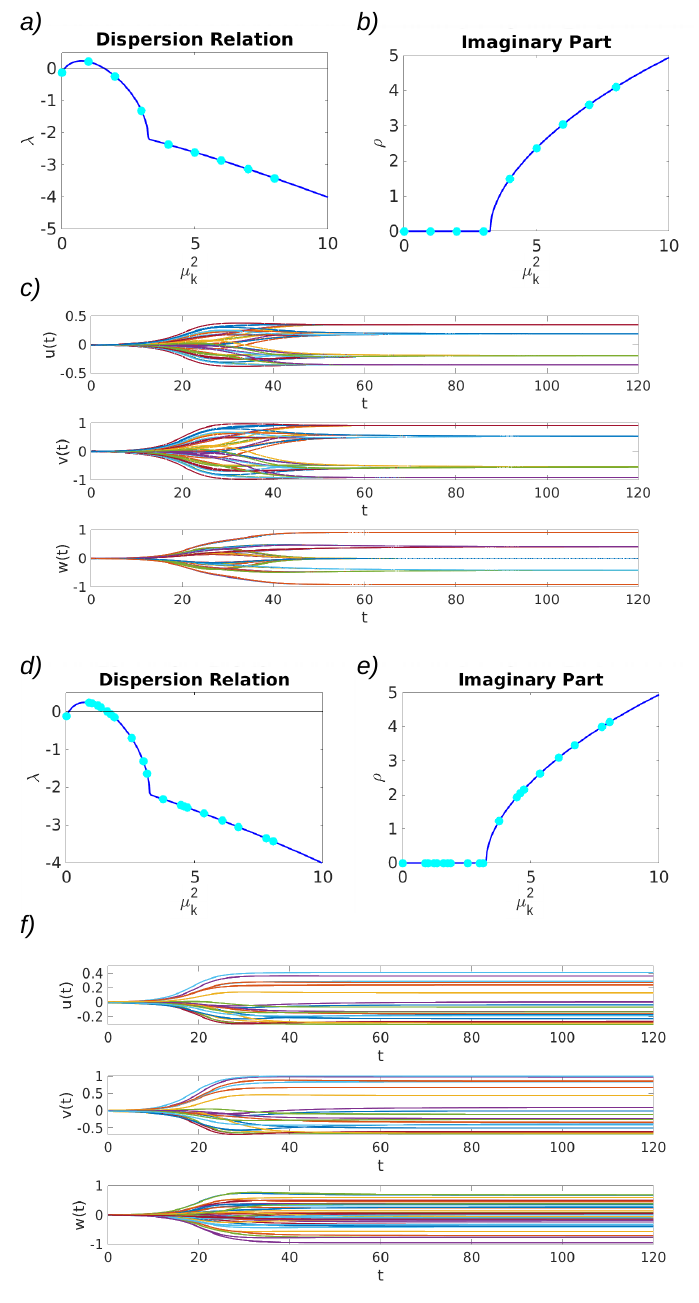}
\caption{\textbf{Dispersion relation and stationary Turing patterns} In panels a) and b) we show the real and imaginary part of the dispersion relation,  respectively, for the case of Turing patterns (i.e., with diffusion) on the $2$D lattice. We can see that, when the real part of the dispersion relation is positive and the corresponding imaginary part is zero, while the modes with non-zero imaginary part have a negative real part. The dynamics of the stationary patterns for the $3$ interacting species $(u,v,w)$ are shown in panel c). Panels d), e) and f) show the analogous setting but for the Random Eulerian Networks. The parameters for both cases are $\sigma_1=0.2,~\sigma_2=-7.8,~\sigma_3=-2,~\eta_1=-3,~\xi_1=-0.5,~\xi_2=5,
~\zeta_1=0.1,~\zeta_2=6,~\zeta_3=6,~\zeta_4=1,~c_1=-1,~c_2=1,~\alpha_u=1,~\alpha_v=1,~\beta_u=1,~\beta_v=0.1$; the initial perturbation is the order of $10^{-2}$.}
\label{fig:stationary}
\end{figure}

Throughout this work, we have always dealt with dynamical patterns, which are made possible by the formalism hereby introduced. However, as we have mentioned in the Main Text, there are cases in which the patterns are stationary, even when one deals with allowing for oscillatory instabilities, such as $3$ interacting species \cite{hata_wave} or hyperbolic reaction-diffusion systems \cite{ZH2016}. \\
For sake of completeness, in Fig. \ref{fig:stationary}, we show, for both the $2$D lattice and the Random Eulerian Network, the dispersion relation and the patterns in a setting in which the latter are stationary, i.e., the parameters are chosen from the green region of Fig. \ref{fig:Tur_Dir_regions}a). We can see that, when the real part of the dispersion relation is positive and the corresponding imaginary part is zero, while the modes with non-zero imaginary part have a negative real part. Hence, the pattern is stationary. \\
Let us conclude by stressing that the fact the imaginary part of the dispersion relation is non-zero in correspondence of the unstable modes does not automatically guarantee a dynamical pattern. In fact, for hyperbolic reaction-diffusion systems on networks, there have been found settings in which the modes responsible for the instability have zero imaginary part, which is non-zero for negative modes, but patterns are oscillatory \cite{jop_carletti}. Also the opposite case can verify, hence, one may have the unstable modes with zero imaginary part, but the patterns can be oscillatory. The latter cases are not common and in general the imaginary part gives a good indication of the resulting pattern, even though one always needs to perform the numerical experiment before claiming the nature of the pattern.

%

\end{document}